\documentclass[twocolumn,dvipsnames]{aastex631}
\usepackage[utf8]{inputenc}

\usepackage{natbib,graphicx,amsmath,amsthm,ulem,color,wasysym}
\usepackage{hyperref,float}
\usepackage[]{xcolor}

\definecolor{darkgreen}{rgb}{0.09, 0.45, 0.27}
\definecolor{darkred}{rgb}{0.90 0.29 0.27}
\definecolor{darkgreen}{rgb}{0.27 0.70 0.50}

\defcitealias{CG97}{CG97}

\newcommand{\y}{\color{red}}

\newcommand{\w}{\color{blue}}
\newcommand{\renu}{\color[rgb]{0.8,0,0.6}}

\newcommand{\be}{\begin{eqnarray}}
\newcommand{\ee}{\end{eqnarray}}
\newcommand{\beqn}{\begin{eqnarray}}
\newcommand{\eeqn}{\end{eqnarray}}
\newcommand{\bi}{\begin{itemize}}
\newcommand{\ei}{\end{itemize}}

\def\refnew#1{(\ref{#1})}

\def\g{\, \rm g}

\def\km{\, \rm km}
\def\cm{\, \rm cm}

\usepackage{ulem}
\newcommand{\x}{\sout}

\newcommand{\bld}[1]{\mbox{\boldmath$#1$\unboldmath}}

\begin{document} 

\title{Repelling Planet pairs by Ping-pong Scattering}

\author[0000-0003-0511-0893]{Yanqin Wu}
\affiliation{Department of Astronomy \& Astrophysics, University of 
Toronto}

\author[0000-0002-1226-3305]{Renu Malhotra}
\affiliation{Lunar and Planetary Laboratory, The University of Arizona}

\author[0000-0003-4450-0528]{Yoram Lithwick}
\affiliation{Dept. of Physics and Asronomy, Northwestern University, 2145 Sheridan Rd., Evanston, IL 60208 \& Center for Interdisciplinary Exploration and Research in Astrophysics (CIERA)}

\begin{abstract}
%While low-mass planets are packed and have nondescript period ratios, there is a key dynamical feature that remains puzzling: the pile-up (and deficit) of planet pairs just outside (inside) of some orbital commensurabilities....
The {\it Kepler} mission reveals a peculiar trough-peak feature in the orbital spacing of close-in planets near mean-motion resonances: a deficit and an excess that are a couple percent to the narrow, respectively wide, of the resonances.
%The {\it Kepler} mission reveals a peculiar feature in the orbital spacing of close-in planets:  a deficit (and an excess) that are a couple percent to the narrow (and the wide) of some mean-motion resonances.
This feature has received two main classes of explanations, one involving eccentricity damping,
%(the so-called 'resonant repulsion'), 
the other scattering with small bodies. 
%\color{orange}
%[should be scattering with small bodies or gas turbulence, as Rein also includes that as possibility]
%\color{black}
Here, we point out a few issues with the damping scenario, and study the scattering scenario in more detail. We elucidate why scattering small bodies tends to repel two planets. As the small bodies random-walk in energy and angular momentum space, they tend to absorb,
%, on average, 
fractionally, more energy than angular momentum.
%{\y does this address your concern?}
This, which we call "ping-pong repulsion", transports angular momentum from the inner to the outer planet and pushes the two planets apart. Such a process, even if ubiquitous, leaves 
identifiable marks only near first-order resonances:  diverging pairs jump across the resonance quickly and %resonance dynamics forces a diverging
%pair to jump across the resonance and 
produce the above asymmetry. % an asymmetric distribution of planet spacings around the  resonance.  
To explain the 
observed positions of the trough-peaks, a total scattering mass of order a few percent of the planet masses is required.
Moreover, if this mass is dominated by a handful of Mercury-sized bodies, one can also explain the planet eccentricities as inferred from transit-time-variations.
%and part of the inclination dispersion within planetary systems.
%We provide a plausible scenario where such conditions naturally arise, and discuss what these implies for the formation of {\it Kepler} planets.. .
Lastly, we suggest how these conditions may have naturally arisen during the late stages of planet formation.
%propose a scenario where such conditions naturally arise. 
%{\w Before these ping-pongs are eventually absorbed into the planets, they are often tidally destroyed and give rise to newer generations of bodies. ... }
\end{abstract}

\section{Introduction}
\label{sec:intro}

%\color{cyan}  [NOTE: In the few places where we have not yet converged, you can feel free to put it however you like, and delete my comment, if you'd like.  I'm satisfied that you've thought about all ofmy points, so it's ok with me if you feel your solution is better.]\color{black}

%{\y to do list: 1) currently, all figure captions are too long, need help shortening; 2) emphasize only 1st order MMR should matter?}

%\color{cyan}
%\noindent
%This is Yoram's pen color.

%It would be nice to come up with a name by which this effect is referred to in the future, and which we can regularly refer to as in this paper. I'm not good at naming. But some possibilities are ping-pong scattering, ping-pong repulsion, repulsion by planetesimals
%{\renu I agree that it would be very effective to name the effect; 'ping-pong scattering' sounds good. }
%\color{black}
%{\w I like 'ping-pong repulsion' slightly better because it contains a sign.} 

%{\y notation: the ping pong's orbital elements are $a$, $e$, $i$, $u$, mass $m$, radius $R$, total mass $m_{\rm tot}$, total number $N$.The planet's mass is $M_p$ (and $M_{p1}$ and $M_{p2}$); radius $R_p$, periods are $P_1$ and $P_2$, and sma is $a_p$.    }

\begin{figure*}
    \centering
 \includegraphics[width=0.95\textwidth]{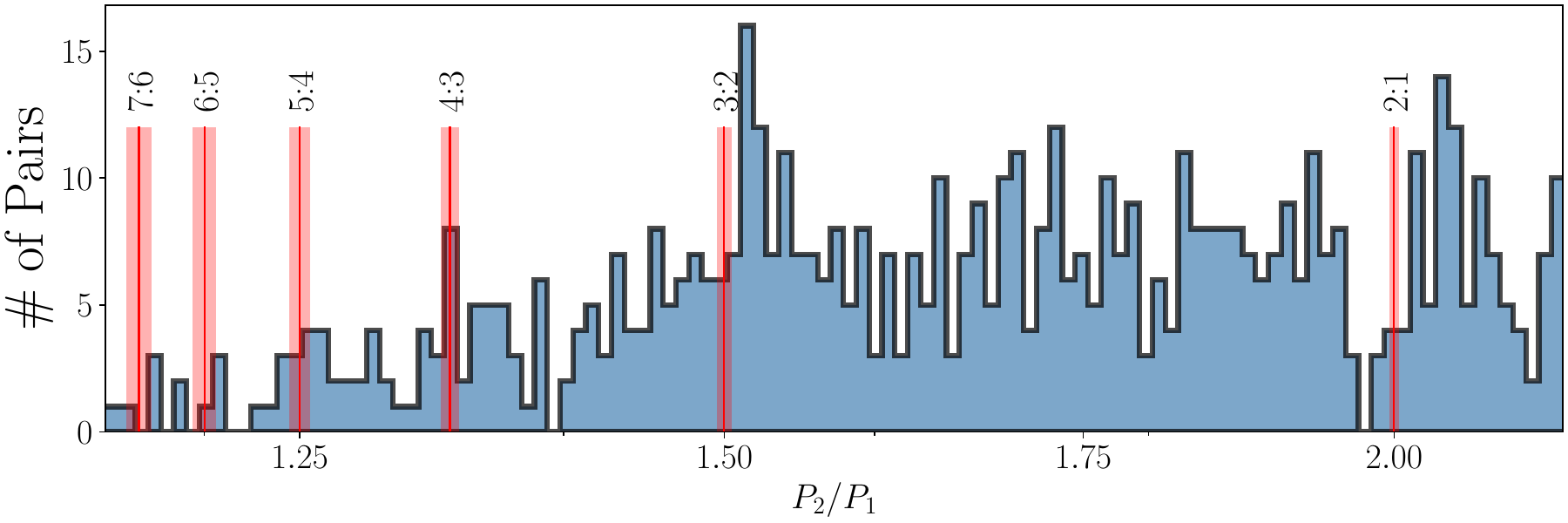}
    \caption{The observed distribution of period ratios for {\it Kepler} planet pairs. The colored lines mark both the centroids and the widths (estimated using eq. \ref{eq:delta}) of  first-order mean-motion resonances.
%    \color{cyan} [could shorten this caption by removing the following] \color{black}
%    We aim to explain the asymmetries around these MMRs, the most significant being the large excess wide of the 3:2 MMR and the clear deficit interior to the 2:1 MMR.
}
\label{fig:dp1}
\end{figure*}

Orbital spacings between planets contain clues to their formation and evolution. 
This information is now abundantly provided by the thousands of multi-planet systems discovered
by the {\it Kepler} mission. 
Most of exo-planets discovered by {\it Kepler} are Earth- to Neptune-sized with orbital periods up to about three years (limited by the lifetime of the {\it Kepler} mission). 

The observed period spacings of neighboring planets have a broad distribution. At the lower end, the distribution appears to be sculpted by dynamical stability, in that spacings of period ratio less than about 1.3 are sparse. Above this, there is a broad peak that extends to period ratio about 2.5, and a long tail of larger spacings. 
Within this generally nondescript distribution, however, there exists one interesting feature. \citet{lissaur,fabrycky} reported a distinct excess of planet pairs just wide of, and a nearly empty gap just narrow of, certain first-order MMRs; Fig. \ref{fig:dp1} illustrates this trough-peak feature. 
According to \citet{stefan}, the statistically most significant amongst these are the large excess
wide of the 3:2 MMR, and the clear deficit  interior to the 2:1 MMR.
These features lie about $1-2\%$ away from exact resonance and have widths that are of the same magnitude. These values are a few times greater than the widths of these resonances (also marked in Fig. \ref{fig:dp1}).

Much work has been expended in deciphering this precious feature.
Theories where planets have undergone some degree of convergent migration predict clustering at the 2:1 and 3:2 MMRs, rather than the trough-peak feature. 
This disagreement could indicate either the absence of convergent migration, or a reorganization post-formation \citep[see, e.g.][]{izidoro2017,ghosh2023}.
There are two main categories of proposals to explain the trough-peak feature:

%see Appendix \ref{sec:app}).
%\color{cyan}
%[How about ``inflating'' the colored vertical lines in Fig. 1 to that predicted width? That would make this point more graphical.]
%\color{black}{\y you can send me the formula for arbitrary MMR width} \color{cyan} I've now checked Eq. A7 in Appendix. 
%It is accurate for 2:1, 3:2, and 4:3.  For 5:4 it looks accurate too, but starting to be perturbed by neighboring resonances.  Note that if you inflate lines, they should be centered on the resonance. (That wasn't obvious to me, but I checked it with REBOUND).
%So, 2:1 should run from 1.995 to 2.005, assuming the standard case of two 8$M_\oplus$ planets.
%\color{black}

%Petrovich, Malhotra & Tremaine 2013; Goldreich & Schlichting 2014; Xie 2014).
\begin{itemize}
\item repulsion by e-damping:
damping of the orbital eccentricities (by friction with gas or planetesimals) leads to orbital divergence of a planet pair,
 akin to the spreading of an accretion disk when its energy (but not angular momentum) is removed by internal dissipation. 
 This effect is strongly amplified near a MMR due to the larger forced eccentricities there, hence the term 'resonant repulsion'
\citep{lithwickwu,batygin}. However, we argue below that updated data do not support this as the dominant mechanism.

\item repulsion by scattering: 
\citet{chatter2015,chatter2022} suggested that, instead of dissipation, scattering of planetesimals can also repel a planet pair and produce the observed feature. Although the repulsion effect is observed in their numerical simulations, there is no physical explanation. Moreover,
the original proposals invoked massive planetesimal belts comparable in mass to those of the planets. 
%claim need primordial resonance pair, but unclear provenance (?)...

\end{itemize}

%$da/dt$: migration/scattering; $de/dt$ dissipation and resonant repulsion....
A few other explanations have also been proposed, including stochastic forcing by turbulence  \citep{rein2012},
finite amplitude libration \citep{goldreich,xie}, and in situ planet mass growth \citep{petrovich}. The first of these is hard to constrain or test, because the disk properties are not well known \citep[see, e.g.,][]{batyginadams}. The other two require
planet-to-star mass ratios near that of Saturn/Sun, much larger than those in most of the Kepler systems, and thus fail to account for the magnitude of the observed asymmetry \citep[e.g., see, \S 5 of][]{goldreich}. 
%{\y Rein says "stochastic could also be from planetesimals", but never proves even its sign, let alone magnitude. So didn't include it as 'scattering' scenario }
%{\renu This is a little bit reminiscent of the famous Goldreich \& Tremaine 1980 paper on disk-satellite torques, although they did argue for large magnitudes but of uncertain sign!}
%system trying to get to the forced states, so some e, so some expansion. total E/J conserved. This is, near the resonance, is of order resonance width and is too narrow to explain the observed feature. There is also the artificialness of a perfectly circular orbit.
%However, we now know that typical masses for these low-mass planets are of order $10 M_\oplus$, making the libration width too low to explain the asymmetry. Also needs special initial condition.

We are motivated to revisit this issue now, a decade after much of the above works, because the past ten years have provided us with some more insights, both observational and theoretical. In particular, the following three considerations have persuaded us that the 'damping' proposal is not the main route. 

\begin{figure}
\centering
\includegraphics[width=0.45\textwidth,trim={0 80 0 0},clip]{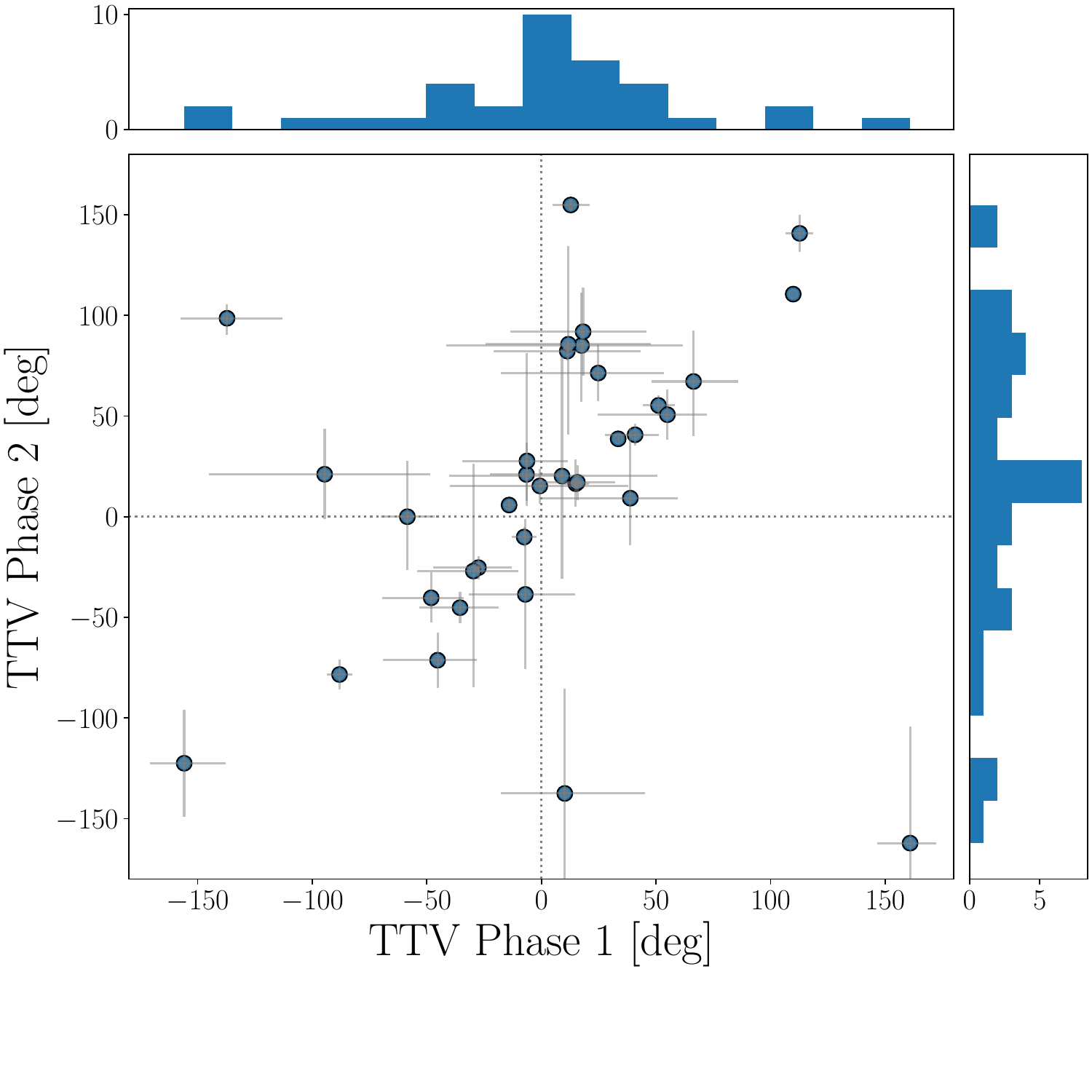}\caption{
%{\w There is a problem with the figure: the x and y errorbars are inverted relative to Sam's figure.  I asked Sam why his table disagreed with his figure, and he said he'd get back to me -- resolved? following write-up correct?} 
%Note also that the distribution of phases predicted by an rms eccentricity is hard to make (it wasn't explained clearly in HL14; it depends on how one models the distribution of $\Delta$'s. However I do it, it will likely not match Sam's). I think we can leave it out without  big loss.
%This is now same as Sam's figure, except restricts to planets where both errorbars are less than 60 degrees, as well as messed-up errorbars.
%
Transit-time-variations (TTV) reveal the non-zero free eccentricities in near-resonant planet pairs, inconsistent with the resonant repulsion. TTV phases (x-axis for the inner planets and y-axis for the outer) for 35 planet pairs around 3:2 and 2:1 MMRs are taken from \citet{Hadden2014}, but shifted by $\pi$ from their convention for the outer ones. The error-bars indicate $1-\sigma$ uncertainties. We have removed pairs with uncertainties greater than $60\deg$, and the mean-error is now $\sim 20\deg$. 
The top and side panels show corresponding histograms for the phases.
Pairs with zero free eccentricities should lie at the origin $(0\deg,0\deg)$. Observed pairs cluster around the origin but with a dispersion that is a few times larger than the mean error, indicating small but non-zero eccentricities.
}
\label{fig:freee}
\end{figure}

\begin{figure}
\centering
\includegraphics[width=0.4\textwidth]{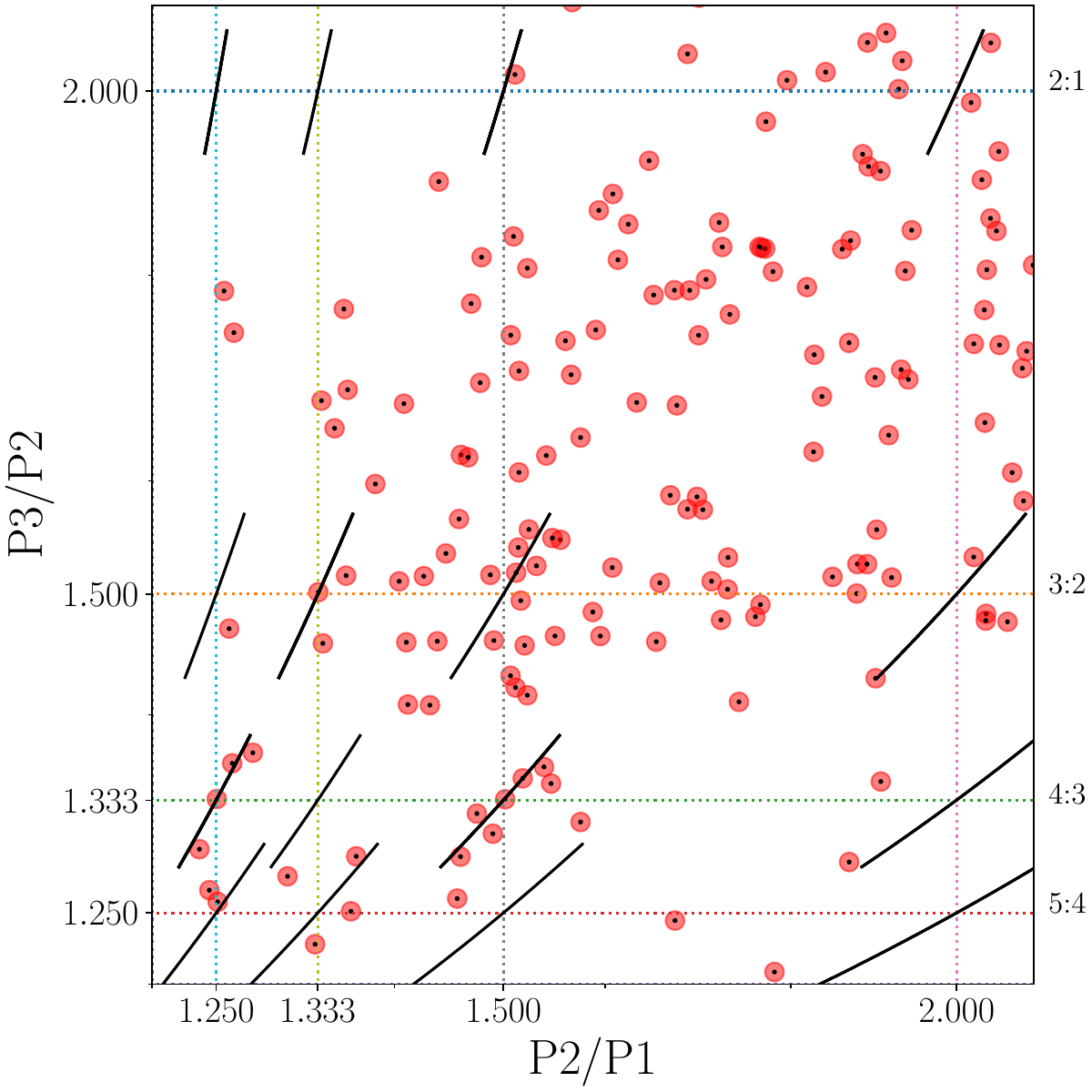}
\caption{If eccentricity-damping is at work, one expects the period ratios of near-resonant triplets to obey the Laplace-like 3-body resonances (black curves, where $k_1 n_1 + k_2 n_2 + k_3 n_3 = 0$ and $k_1 + k_2 + k_3=0$, limited to $|k_1|+|k_2|+|k_3| \leq 12$). This does not seem overwhelmingly common.
}
%shows is it not damping, study by \citet{charalambous2023}misleading, looking under the lamp-post }
\label{fig:pratio-3body}
\end{figure}

First, to produce the observed asymmetry  ($\sim 1\%$ in distance from a MMR) via damping, an extensive period of damping (over 100 $\tau_e$, where $\tau_e$ is the e-damping time) is required \citep{lithwickwu}. One then expects all near-resonant pairs to have vanishingly small free eccentricities (eccentricities on top of that forced by planets).
If so, all transit-time-variations (TTVs) should have transit phases pinned at some special values. This is not consistent with data (Fig. \ref{fig:freee}). Instead, TTV measurements returned free-eccentricities of order a couple percent \citep{lxw,wulithwick,Hadden2014,Haddden2017}.
%[Note also: I can't directly show individual free e's in Fig 2. That can only be done from the data in HL17, not HL14.]
\citet{choksi} attempted to remove this tension by invoking a third planet in the system that is eccentric. However, even though neighbouring companions are indeed quite commonly observed, its being eccentric will be surprising -- the extensive e-damping suffered by the planet pair would have to, somehow, spare the companion.\footnote{Giant companions at larger separations,  an uncommon occurrence, may be more plausible for this scenario.}

The same e-damping epoch would also have flattened the pairs down to co-planar orbits. However, while mutual inclinations  in transiting systems are small, they are distinctly non-zero.  
\citet{fabrycky} estimated an inclination dispersion of $1-2\deg$, based on the impact parameters. For systems with high-multiple transiting planets, \citet{zhuwu} inferred a somewhat smaller value of $\sim 0.8\deg$.

Lastly, the triplet structure appear  inconsistent with extensive e-damping. \citet{batygin} considered a chain of 3 planets subject to e-damping and found that they should be  deposited along the so-called Laplace-like 3-body resonances ($k_1 n_1 + k_2 n_2 + k_3 n_3 = 0$, with $k_1+k_2+k_3=0$ and $n_i$ the mean-motions).
 %{\w as the planets diverge from each other.} {\y I concur with Yoram's point that it takes migration time, not damping time.}
%\color{cyan}
%The timescale is that to migrate the planet to the resonance by resonant repulsion.
%\color{cyan}
%[I'm finding this in my sims.  Maybe you're thinking about different initial conditions than I'm using?  i.e., where the i.c. is close to your black curves in Fig 4, but differing in $e$.]
%\color{black}
With such a combination of orbital periods,  forced eccentricities on the middle planet from its two neighbours cancel effectively, because the they have the same forcing periods but opposite phases. 
So these 'three body resonances', different from the usual mean-motion resonances, minimize the forced eccentricities.  However, such a preference does not seem overwhelmingly common in known triplets \citep[][also see Fig. \ref{fig:pratio-3body} here]{pichierri}, though a more thorough statistical study on the degree of prevalence is merited  
%{\y is the above toning down sufficient now?}
%\color{cyan}
%[Works for me.  I commented out  the old comments that were here]\color{black}
%[I'm not sure how to proceed regarding Fig 4. It appears to me that the circles in the figure  show some clustering around black curves. And my damping sims also show that, ``much of the time'', a damped 3-planet system also tracks the black curves (considering only the (1.5,1.5) crossing thus far). This seems to merit a more extended 
%investigation, perhaps in a section at the end (?).  But I'm not sure both of you agree with me that Fig 4 suggests consistency with both resonant repulsion and, possibly, ping-pong repulsion too.]
\color{black}
%{\renu Without some formal statistical analysis, it seems to me that the clusterings in Fig 4 are of marginal significance; my suggestion is to not belabor this but simply state that the triplet clusters may be consistent with both resonant repulsion and with ping-pong repulson, and merit a closer investigation in a future study. }

%\color{cyan}
% From your new figure 4, it looks like some resonance crossings do show evidence for 3br especially (1.25,1.33) and (1.5,1.33) and less so (1.5,1.5).  Maybe visual bias; or maybe could happen also with pingpongs.  
% Also, I did some experiments with e-dampings. 
% Most do end up along your curve segments in Fig 4, though there are some exceptions.  
% So I think can stick with Fig 4.  
 %{\renu Please see my suggestion at the end of the previous para.}
%\color{black}

%\bigskip
These failures of the e-damping proposal motivate us to take a deeper look at the 'scattering' proposal. We would like to understand the physical reason %\x{for orbital repulsion with ping-pong scattering} 
why scattering leads to orbital repulsion, to assess the amount of scattering mass required to produce the observed asymmetry, and to determine whether the scattering process can self-consistently explain the observed free eccentricities and free inclinations. 

In this work, we  use the term "ping-pong repulsion" to refer to the process in which planetesimals ("ping-pongs") scatter off of a pair of planets and cause the planets to repel each other. We explain the underlying cause for repulsion (\S \ref{sec:dissect}), and clarify the
conditions required to explain the observed asymmetry %features near the first order MMRs in the {\it Kepler} systems 
(\S
\ref{sec:requirements}). We end by suggesting a plausible scenario where these conditions naturally arise, as well as other extant issues (\S \ref{sec:discussion}). 

\section{Ping-Pong Repulsion Dissected}
\label{sec:dissect}

It is known that planets can undergo divergent migration when scattering
planetesimals. For example, Jupiter can eject bodies that are
sent inward by Neptune, so Jupiter loses energy while Neptune
gains \citep{FernandezIp,Hahn1999}. 
This well-known mechanism is, however,
different from our problem here.
One significant difference is that Jupiter easily ejects bodies from the Solar System because its surface escape speed exceeds its orbital speed. 

By contrast, {\it Kepler} planets are
in the opposite limit. They lie so deeply in their stars' gravitational potential well, they cannot easily eject bodies. Instead, they relentlessly scatter
the planetesimals 
%Jupiter, with a Safronov number of %$\Theta \sim 10$, is efficient at ejecting small bodies. In contrast, }
%{\w and have small Safronov numbers ($\Theta \ll 1$).} \x{Their surface escape velocities fall much below the orbital escape values.} As a result, a planetesimal \x{(which we call ping-pong from now on)} can't be easily ejected but \x{has to be} {\w is instead} 
until physical collision and merger occur.  In this case, do the orbits of a pair of scattering planets pull apart or converge?
%{\renu [I would guess that merger with the star is the most common final outcome ($t\longrightarrow \infty)$, and the ejection very infrequent - even neglible - if surface escape velocity of the planets is less than $(\sqrt2-1)v_{orb}$]}

\subsection{Ping-pongs repel a pair}
\label{subsec:repel}

Numerical experiments often show that, when the planets scatter small bodies,  divergent migration takes
place \citep{chatter2015,kratter2018,chatter2022}.  Here we demonstrate this effect and provide a physical explanation for the repulsion:  an uneven phase space.

\begin{figure}
    \centering
    \includegraphics[width=0.45\textwidth]{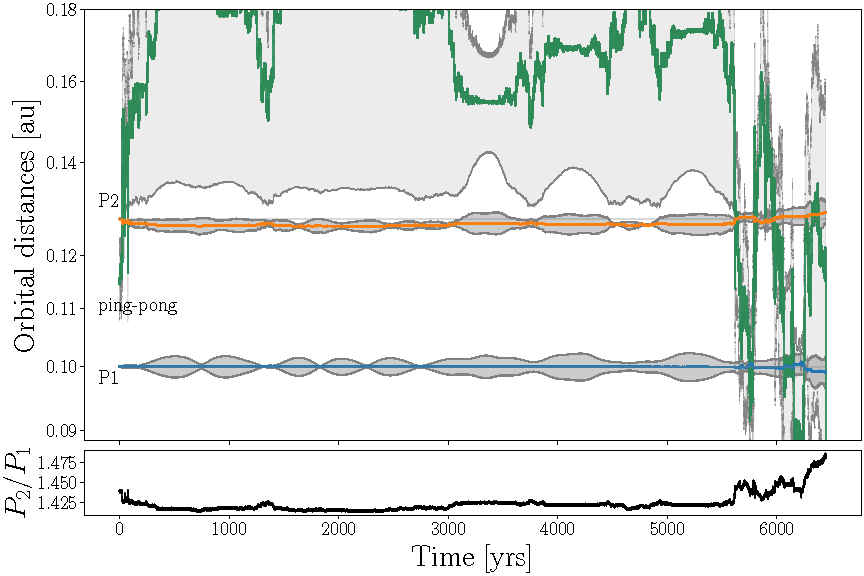}
    \caption{
    %A pair of planets scattering and finally absorbing one ping-pong. The top panel shows the orbital extent of each: the colored curves indicating the semi-major axis of all bodies, and the shaded regions around them encompassing from periaps to apoaps.     The two planets each has $8M_\oplus$ and the ping-pong mass is $1.3\%$ of the combined planet mass (or, $\sim 4$ Mercury masses). It repels the pair from an initial period ratio (shown in the lower panel) of $P_2/P_1 = 1.44$ to the final $1.48$, or a change of $\sim 3\%$.
   % $2 M_\male$ 
      %50 pluto = 1 mas; 100 pluto = 4 mercury
      %Planet eccentricities and inclinations are also excited during the process. The final absorption of the ping-pong by one of the planets (at $\sim 6500$ yrs) does not noticeably affect the outcome.
%      \color{orange}
%[why not extend timescale for which you show planets' orbits, until end of plot, so we can see that absorption of ping-pong doesn't affect the orbits?]
%      \color{black}
   A pair of planets scattering and finally accreting a ping-pong. 
   The planets are each $8M_\oplus$,  and the ping-pong  is $1.3\%$ of the combined planet mass. 
   The top panel shows
   the
   semi-major axis, apoapse, and periapse for each body.
   The bottom panel shows the planets' period ratio, which increases by $\sim 3\%$ by the time the ping-pong is accreted.
    The final accretion has little effect on the period ratio.
      }
\label{fig:neptune3}
\end{figure}

Consider first the numerical example of one ping-pong (Fig. \ref{fig:neptune3}). It is in an
initially nearly circular ($e=0.05$), nearly co-planar
orbit between two circular, co-planar
planets (the inner planet at $0.1$au). It is
excited into orbit crossing with the planets by dynamical instability.\footnote{A more circular ping-pong may sometimes not reach this state.}
%Let planets be placed at $a_{\rm p1}$, $a_{\rm p2}$, with equal mass $M_p$ and circular/coplanar orbits; the planetesimal has mass $m_p$ and at orbit $a,e,i (=0)$. 
Subsequent close-encounters force the ping-pong to random-walk in energy and angular momentum space, until it  is accreted by one of the planets. Even though the ping-pong gains or loses energy with roughly equal
probability during each encounter, 
by the time of the merger, the ping-pong has successfully transferred energy and angular momentum from the inner to the outer planet and pushed them apart. 
%{\w (and MMR?)},\color{cyan} Could phrase as: ``During any conjunction,'' which encompasses both close-encounters and MMR's, but approximately.   Or: During any interaction\color{black}
Such an outcome appears common among most simulations.

We numerically integrate the system using the REBOUND code \citep{rebound2012} with the IAS15  integrator \citep{ias15} at a minimum timestep of $10^{-10}$ day. Masses for the planets and the ping-pong are $M_p$ ($ = M_{p1} = M_{p2}$) and $m$, respectively. Here we take $M_{p} = 8 M_\oplus$ and $m = 1.3\% \times (M_{p1}+M_{p2}) \approx 4$ Mercury masses. We assume the ping-pong has a negligible radius ($R=0$), while the planets have radii $R_p = 1.5 R_\oplus$. The merger occurs after a few thousand years and is modelled as conserving linear momentum but dissipating all relative motion.  
%\color{orange}
%[all or most?  I thought most. eg, if it climbs high on a jacobi curve exterior to the two planets, then it can push them together.]
%\color{black}

%\clearpage

\begin{figure}
   \centering
    \includegraphics[width=0.46\textwidth]{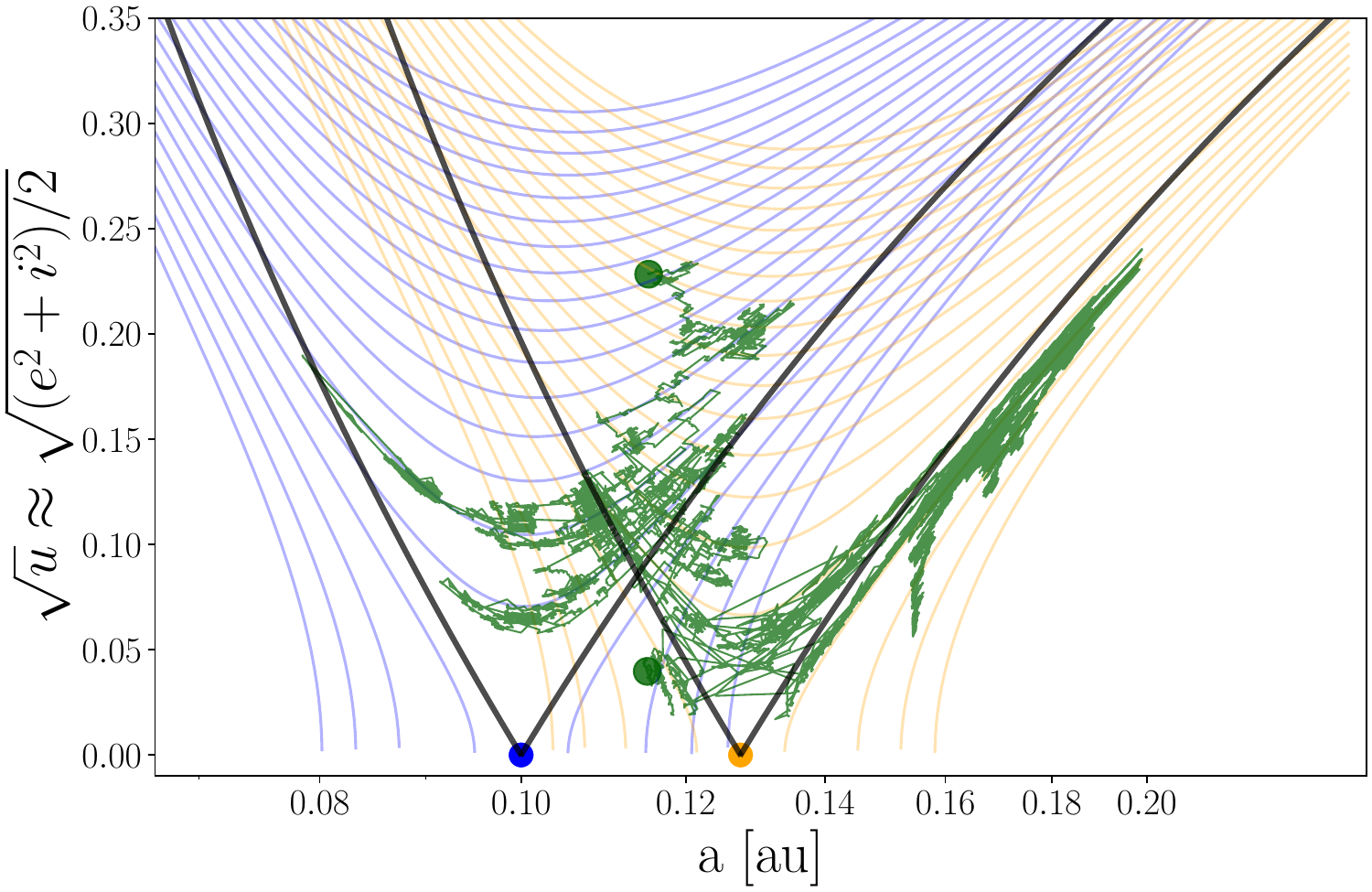}
    \caption{   
%    The detailed dynamics for the ping-pong from Fig. \ref{fig:neptune3}.     Gravitational jostling by the planet pairs forces the ping-pong to random-walk (green jagged trajectory) in the plane of semi-major axis (horizontal axis) and $u$ (vertical axis), where $u$ describes the ping-pong's random speed (or angular momentum deficit, eq. \ref{eq:defineu}). It starts from the green circle at lower u and ends at the one at higher u, when it merges with a planet. The two planets are marked by the blue and orange circles, and the same colors mark their respective curves of constant Jacobi integral. The two sets of thick
%For an initially circular ping-pong at $0.115$au,
% black curves roughly delineate the regions within which  orbit-crossing with either of them is  possible (assuming $i=0$). At any given time, the ping-pong primarily interacts with only one of the planets so it follows one of the Jacobi curves. It gradually diffuses upward along as it traverses the different Jacobi curves. 
 % 
    Ping-pong dynamics in the $a$-$\sqrt{u}$ plane, where $u$ is random motion. The green
    jagged trajectory traces the ping-pong from Fig. \ref{fig:neptune3}, with its initial and final states
     shown as the lower and upper green circles.
    The blue and orange circles represent the planets.  
    Each planet's 
   curves of constant Jacobi integral are shown with its corresponding color.  The ping-pong's trajectory alternates between
    tracking the blue and orange  curves, as it diffuses upwards.
    The black curves show the regions within which the ping-pong
    can cross a planet's orbit (for $i=0$).
          }
\label{fig:x_jacobi}
\end{figure}

\begin{figure}
    \centering
    \includegraphics[width=0.46\textwidth]{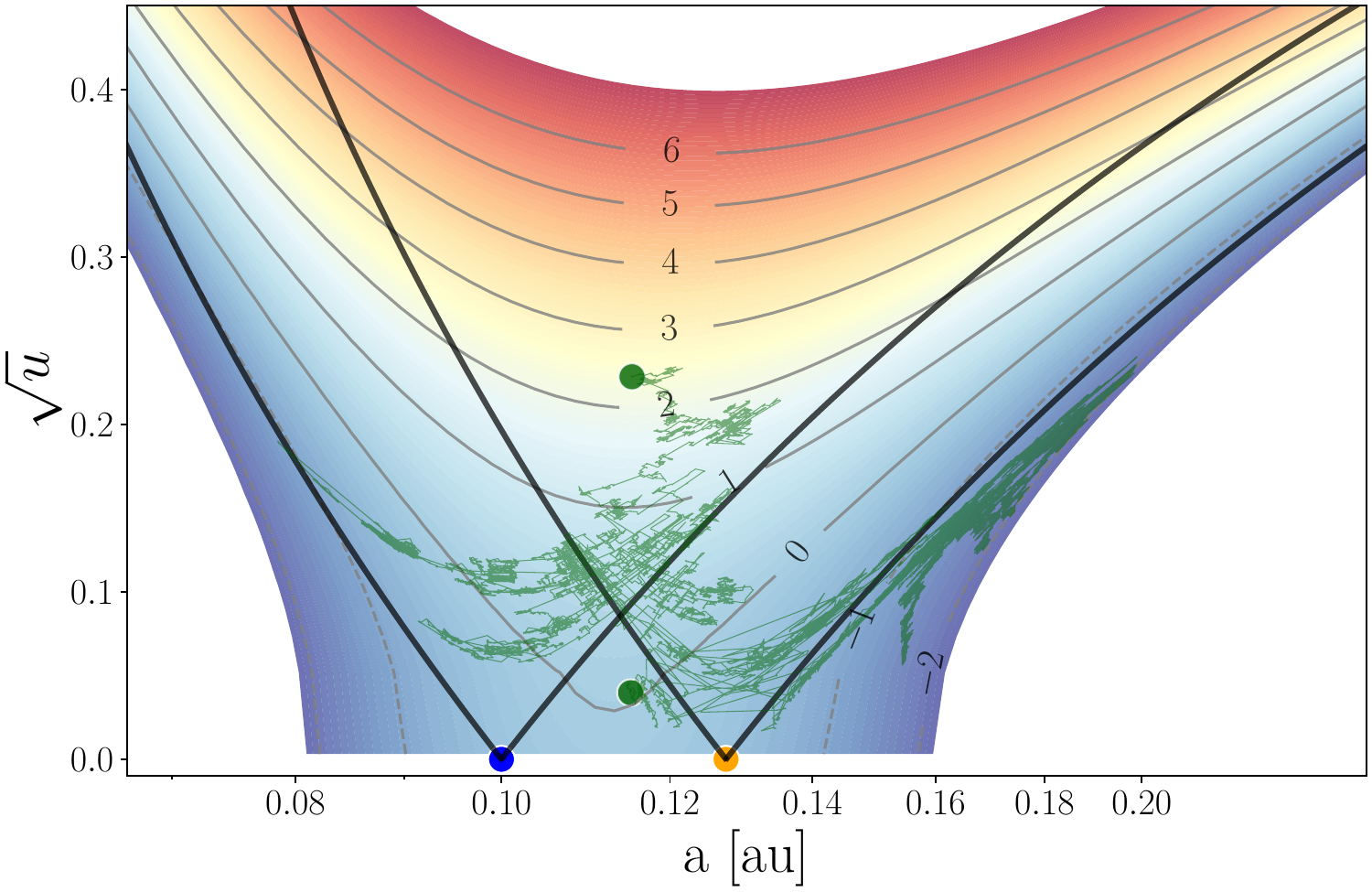}
    \caption{Origin of the repulsion.  Again for the case in Figs. \ref{fig:neptune3}-\ref{fig:x_jacobi}. Here, instead of the Jacobi curves, the background colours and contours indicate the amount of planet repulsion ($\Delta (P_2/P_1)$), measured in unit of $\epsilon=m/(M_{p1}+M_{p2})$, as the ping-pong random walks in the a-u space. 
      %also corresponds to the range of mobility for the ping-pong. 
      Most of the phase space that the ping-pong can access leads to repulsion. Orbital change due to ping-pong accretion is not included here.}
\label{fig:x_dp5}
\end{figure}

To understand this average fate of repulsion, we turn to concepts like Jacobi integral and phase space.
When the ping-pong interacts with a single planet, it very nearly conserves its
Jacobi constant relative to that planet
\footnote{The conservation of Jacobi constant is violated by a correction
that is $O(m^2/M_{p}^2)$. 
Note that our definition of the Jacobi constant differs from the usual one by an unimportant overall constant.}
\begin{equation}
{\rm Jacobi}  = \sqrt{a}(1-u) + {n\sqrt{a}\over 2n_{p}}
\label{eq:jacobi}
\end{equation}
where $n=\sqrt{GM_*}a^{-3/2}$ is the ping-pong's mean-motion, 
$n_{p}$ that for the planet, and 
\be
 u \equiv 1-\sqrt{1-e^2}\cos i\approx (e^2+i^2)/2 \ ,
 \label{eq:defineu}
\ee
%describes the ping-pong's random motion.
is related to the angular-momentum-deficit \citep[AMD,][]{laskar1997} via 
AMD=$m\sqrt{a}\,u$. Curves of constant Jacobi values, for individual planets, are plotted in Fig. \ref{fig:x_jacobi}. At any given time, the ping-pong chiefly interacts with, and
%one of the planets and follows in scattering from that planet it 
evolves along a curve of constant Jacobi with respect to, that planet.
% Basically, when the ping-pong is qon planet 1's Jacobi curve, the change in its energy and angular momentum are in the proportion $n_1$ (mean motion of planet 1).  As a result, if planet 1 pushes the p-p outward, the planet loses energy and angular momentum in the proportion $n_1$, and therefore it moves inward while remaining on a circular orbit.
%It pushes the planet in the direction opposite to its own.  
But when it switches its main scatterer, the ping-pong can switch Jacobi curves. This switching-over allows the ping-pong to random-walk through a large phase space. Over time, the ping-pong appears to climb the Jacobi ladder to ever higher values of $u$ 
%{\y Renu: higher values of $u$ (lower values of the Jacobi integral).?}
because the phase space volume at high eccentricity and high inclination  is larger. 

Fig. \ref{fig:x_dp5} presents  the consequences of such a random-walk: repulsion. Regardless of the ping-pong's mass, the following conservation of total energy and angular momentum apply before the merger event,
%{\y Now, let's relax the assumption of a massless ping-pong. A scatterings of a small mass ping-pong has a back-reaction on the scatterer.}
%Before the ping-pong is accreted, the total energy and angular momentum of the system are conserved:
\begin{eqnarray}
\sum_{i=1}^{2} {{M_{pi}}\over {a_i}}+ {m\over{a}} & =& {\rm const}\, ,\nonumber \\
\sum_{i=1}^{2} M_{pi} \sqrt{a_i} (1-u_i) +
m \sqrt{a} (1-u) & = & {\rm const}\, .
\label{eq:conservation}
\end{eqnarray}
If we simplify by setting $u_1 = u_2 = 0$, we can predict the change in planet spacing (measured as changes in the period ratio, $\Delta (P_2/P_1)$) as a function of $(a,u)$. This is shown by the colour contours in Fig. \ref{fig:x_dp5}. As the ping-pong climbs the Jacobi ladder, it tends to increase in $u$ (angular momentum deficit). So fractionally speaking, it removes more energy than angular momentum from the planet duo. This repels the pair, increasing in degree as the ping-pong diffuses to higher values of $u$. There is an analogy of this physics in an accretion disk: as kinetic energy is converted into heat by, e.g., turbulence, the conservation of angular momentum dictate that the disk must spread radially.
Another analogy, less accurate, is when a real ping-pong push apart two paddles as it scatters off of them.
 %\color{cyan}
%[The analogy works better if you talk about angular momentum and momentum here, rather than energy. Because the ping-pong doesn't transfer energy.]
%\color{black}

In summary, driven by the urge to explore more phase space, the ping-pong on average repels the planet pair. 

In the meantime, the scatterings also affect
%\color{cyan} [excite $\rightarrow$ affect, as it also damps planet e if it is high]  \color{black}
the eccentricities and inclinations of the planets. This can be thought of as a process of AMD equipartition  \citep{wetherill1989}. This yields, to order of magnitude, $e_p \sim \sqrt{m/M_p} e$, and inclinations that are about half the eccentricities \citep[see][and Fig. \ref{fig:comparison} here]{Kokubo1998}. 
%Moreover, the planets have finite sizes and can accrete the ping-pong. In the following, we model the acccretion process. We also consider the dynamics that sets in when the planet pair cross a first-order resonance. 

\subsection{Accretion and the Average Repulsion}

%time to scatter to unity eccentricity is, assuming random-walk with
%stepsize $v_{\rm hill}$ in every hill entry time,
%\begin{equation}
%t_{\rm scatter} \sim \left({{v_{\rm kep}}\over{v_{\rm hill}}}\right)^2 t_{\rm hill} \sim%\end{equation}
%where $t_{\rm hill} \sim P_{\rm orb} (a_p/R_{\rm hill})^2$
%{\w REBOUND e rises from 0.05 to 0.1 by 80 yrs; to 0.15 by 840yrs; to 0.2 by 4500yrs; }

The planets have finite sizes, so the ping-pong will be accreted after a time.  This has two effects: it truncates the time for  the  ping-pong to transfer energy; second, the inelastic collision modifies the planetary orbit.

\begin{figure*}
    \centering
    \includegraphics[width=0.98\textwidth]{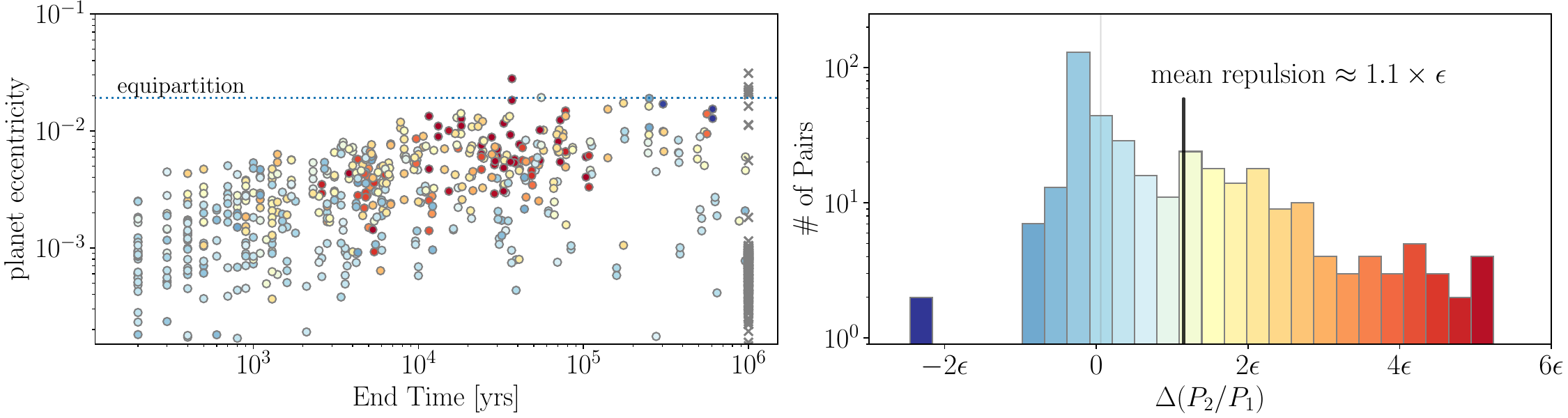}
    \caption{
%Ensemble results by repeating the simulation in Fig. \ref{fig:neptune3} for 1000 times, but with different orbital phase angles for the ping-pong.The left panel records the final planet eccentricities at the time of merger (x-axis), with colors corresponding to the amount of orbital repulsion (see the right panel).The mean survival time for the ping-pong is $\sim 10^4$ yrs. Systems in which the ping-pong survives longer, and can diffuse to larger eccentricities andinclinations, tend to experience larger repulsion, and the planet eccentricities  tend to approach the equipartition value ($\sim (m/M_p)^{1/2}\, e$, the dotted line evaluates for $e = 0.3$). 
%      Ping-pongs that survive beyond $\sim 10^4$ yrs can bring their perturbers to equi-partition. 
%by which equipartition should have been reached.
%About $1/3$ of the systems (marked as gray crosses) remain in  limbo even after the integration limit of $10^6$ yrs.  They are discarded from the analysis.  
        %Many ping-pongs were accreted soon after orbit-crossing and are ineffective at repelling (the spike at $\Delta (P_2/P_1)=0$). 
% The right panel shows the distribution of final repulsion, measured after the inelastic merger. Most pairs ($\sim 85\%$) experience repulsion, with the mean $\Delta (P_2/P_1)$ being of order the mass ratio, $\epsilon \equiv m/(M_{p1}+M_{p2})$.These experiments do not involve a MMR crossing. 
%pairs entering the shaded region gains an extra push by the 3:2 resonance and also acquire a bonus eccentricity.     
    Ensemble results from 1000 simulations that are identical to 
    Fig. \ref{fig:neptune3}, except with randomized initial orbital
    phases for the ping-pong. The left panel shows the final planet
    $e$ vs. time of merger.  The right panel shows the 
    histogram for the net amount of pushing,   for the ensemble. The  color code in both panels quantifies the amount of repulsion.  
    The left panel also shows that in systems where the ping-pong
    lives longer, the planets approach closer to equipartition eccentricity ($\sim (m/M_p)^{1/2}\,
       e$, the dotted line evaluates for $e = 0.3$).
       Around $1/3$ of the systems (marked as gray crosses) remain in  limbo even after the integration limit of $10^6$ yrs.  They are discarded from the analysis.  
 %      [IT MAY be more logical to swap the two panels, as the right panel is the main result.]
}
    \label{fig:dp0}
\end{figure*}

The accretion time, for an orbit-crossing ping-pong, is (ignoring gravitational focussing),
\begin{equation}
    t_{\rm merger} \sim {1\over 4} P_{\rm orb} \times {{2 \pi a^2 }\over{\pi
        R_p^2}} \sim 4 \times 10^4  {\rm yrs}\,
    \left({{a}\over{0.1 {\rm au}}}\right)^{3.5}\,
    \left({{R_p}\over{1.5 R_\oplus}}\right)^{-2}
   \, .
    \label{eq:tmerger}
  \end{equation}
% where we have adopted a planet radius of $1.5 R_\oplus$. 
%{\w REBOUND 50/100 by 3400 yrs; 40/100 by 7800yrs; 37/100 by $10^4$yrs. }
This estimate is confirmed by numerical simulations (Fig. \ref{fig:dp0}), 
%The mean life-time of ping-pongs that are excited to orbit-crossing is $\sim 7\times 10^4$ yrs (eq. \refnew{eq:txsmerger}).
with the caveat that gravitational focussing may be important early-on, before the ping-pong is dynamically excited. %in some simulations, the ping-pong is accreted onto the planet rapidly, due to the effect of gravitational focussing before dynamical excitation.  
%{\y note for renu: compare with Morrison-2015, who show that the accretion of massless particles (on near-circular orbits) in the vicinity of a single planet is on timescale $10^4$ orbital periods; this is shorter, likely due to greater gravitational focussing of near-circular initial particle orbits}
 
%Fig. \ref{fig:dp5} displays the fate of different ping-pongs. 
% actual number 7e4 for 100 pluto mass sim; 5e4 for 10 pluto mass sim
A ping-pong that survives longer can diffuse to higher $u$ (AMD) and can cause more repulsion. As this is a stochastic process, we gauge the average amount of repulsion by simulating  the ping-pong with randomized initial phases 
(Fig. \ref{fig:neptune3}). 
%but with randomized phases for the ping-pong's orbital elements.  
Summing over these results, we find an average repulsion of
\begin{eqnarray}
\Delta \left({{P_2}\over{P_1}}\right) & \equiv &
\left({{P_2}\over{P_1}}\right)_{\rm final} -
\left({{P_2}\over{P_1}}\right)_{\rm init} \nonumber \\
&\approx & 1.1\times {{m}\over{M_{p1}+M_{p2}}} = 1.1 \times \epsilon\, ,
\label{eq:dpp}
\end{eqnarray}
where $\epsilon$ is the fractional mass of the ping-pong. This is also the result one obtains if the single ping-pong is split into an infinite number each with an infinitesimal mass, so we call this result the 'continuum limit'. 
The factor $1.1$ is valid for our set of parameters, i.e.,  near 3:2 MMR,\footnote{The value  is slightly less for 2:1 MMR. %{\w haven't sorted out how large it is}.
} and with our assumed accretion cross-section. If the planets have larger accretion cross sections, the survival time is shorter and the
repulsion will be correspondingly smaller.
%However, we argue in \S\ref{subsec:???}  that a tenuous hydrogen envelope is not relevant for accretion.

The mergers themselves typically cause negligible changes to the planet spacing. Consider a planet on a circular orbit that merges with a
ping-pong. The change in the planet's semi-major axis ($\delta a_p$) is given by the change in
its specific energy, or
$\delta a_p/a_p\approx {2m\over M_p}{\bld{V\cdot\bld{\Delta v}}/ v^2}$,
where $\bld{V}$ is the planet's initial (circular) velocity, and
$\bld{\Delta v}$ is the ping-pong's pre-collision velocity relative to the
planet. Taking the guiding center approximation, the magnitude 
$|\bld{\Delta v}|/V\sim e$, where $e$ is the ping-pong's pre-collision
eccentricity.  
%And the average value of $\bld{\Delta v}$ vanishes. 
Therefore every time a ping-pong merges with the planet,
it changes the planet's $a_p$  by $|\delta a_p/a_p|\sim  (m/M_p)\, e \sim \epsilon \, e$. This is smaller than that in eq. \refnew{eq:dpp} by a factor of $e$. In addition, this change has random signs, because the angle between $\bld{\Delta} v$ and $\bld{V}$ is random. So we can typically ignore this effect.

%Another way to argue about the above result is that a merger event dissipates total energy -- in the case of a perfectly inelastic accretion, all relative energy in the centre-of-mass frame is lost. However, a merger event conserves the total orbital angular momentum almost perfectly, because by definition an accretion event has a small impact parameter and removes little angular momentum (into planet spin). So accretion typically leads to increase of planet spacing.
%[I don't understand argument in second paragraph. Is it based on the fact that if you start off with two planets on nearly circular orbits, and they lose energy at constant angular momentum, then they must spread? My problem is that I'm not sure how that fact changes when you have two planets plus a ping-pong.  You could say that the ping-pong is just a conduit for transferring energy and angular momentum. But part of that 'conduit'ing'  is being used for spreading the two planets pre-collision ]

\subsection{Boost when Crossing a Resonance}
\label{subsec:boost}

The above picture of gradual repulsion is modified when the planet pair encounters a first-order MMR. At the resonance separatrix, a rapid and coherent  energy and angular momentum transfer between the pair ensues.  
The pair pushes each other apart quickly,  on the timescale of resonance libration.

The dynamics of a divergent resonance crossing is summarized in 
Appendix \ref{sec:app}.
%Fig. \ref{fig:mmr}.
%\color{orange} [what figure are you referring to?]  \color{black}
%In Appendix \ref{appendix:??},
The amount of repulsion contributed by the 3:2 MMR is, 
\begin{equation}
\Delta \left({{P_2}\over{P_1}}\right)_{3:2\rm MMR}  
 \approx 0.0095 \left( {M_p\over 8M_\oplus} \right)^{2/3}
%\hskip0.5in 
%{\it when}  \left({{P_2}\over{P_1} \right)_{\rm init} + \Delta \left({{P_2}\over{P_1}}\right) < |1.5 - \delta/2|
    \label{eq:dpp2}
\end{equation}
%where $\delta \approx$ {\y $\mu^{2/3}?$} $0.01$ is the resonance width for 3:2 MMR. It is $???$ for 2:1 (Appendix \ref{appendix:??}. 
for equal-mass planets.  
For the 2:1 resonance, the numerical coefficient is instead $0.0083$.
%{\y oddly, Chatterjee \& Ford said libration width in 2:1 is $78.43 M_P/M_\odot$, where is that from?} \color{orange}
%[Chatterjee \& Ford are wrong. Though numerically, their answer is similar enough to ours not to matter too much.]
%[I'm assuming here that you really want $\Delta(P_2/P_1)$, and those are the numbers I'm quoting for. But, are you sure it's not $\Delta(P_2/P_1)/(P_2/P_1)$ that you really want?  eg, in Eq. 4? ]
%\color{black}
Appendix \ref{sec:app} also provides an estimate for the associated eccentricity kick.  We illustrate these results in Fig. \ref{fig:simplepush}, where we numerically simulate a pair of planets gently pushed apart to cross the 3:2MMR.

This resonance boost accelerates the repulsion at the MMR. It leads naturally to a deficit of pairs to the narrow of the MMR, and to an excess to the wide.  Together with the ordinary repulsion (eq. \ref{eq:dpp}), one can then account quantitatively for the {\it Kepler} 
asymmetry.

\begin{figure}
    \centering
 \includegraphics[width=0.49\textwidth]{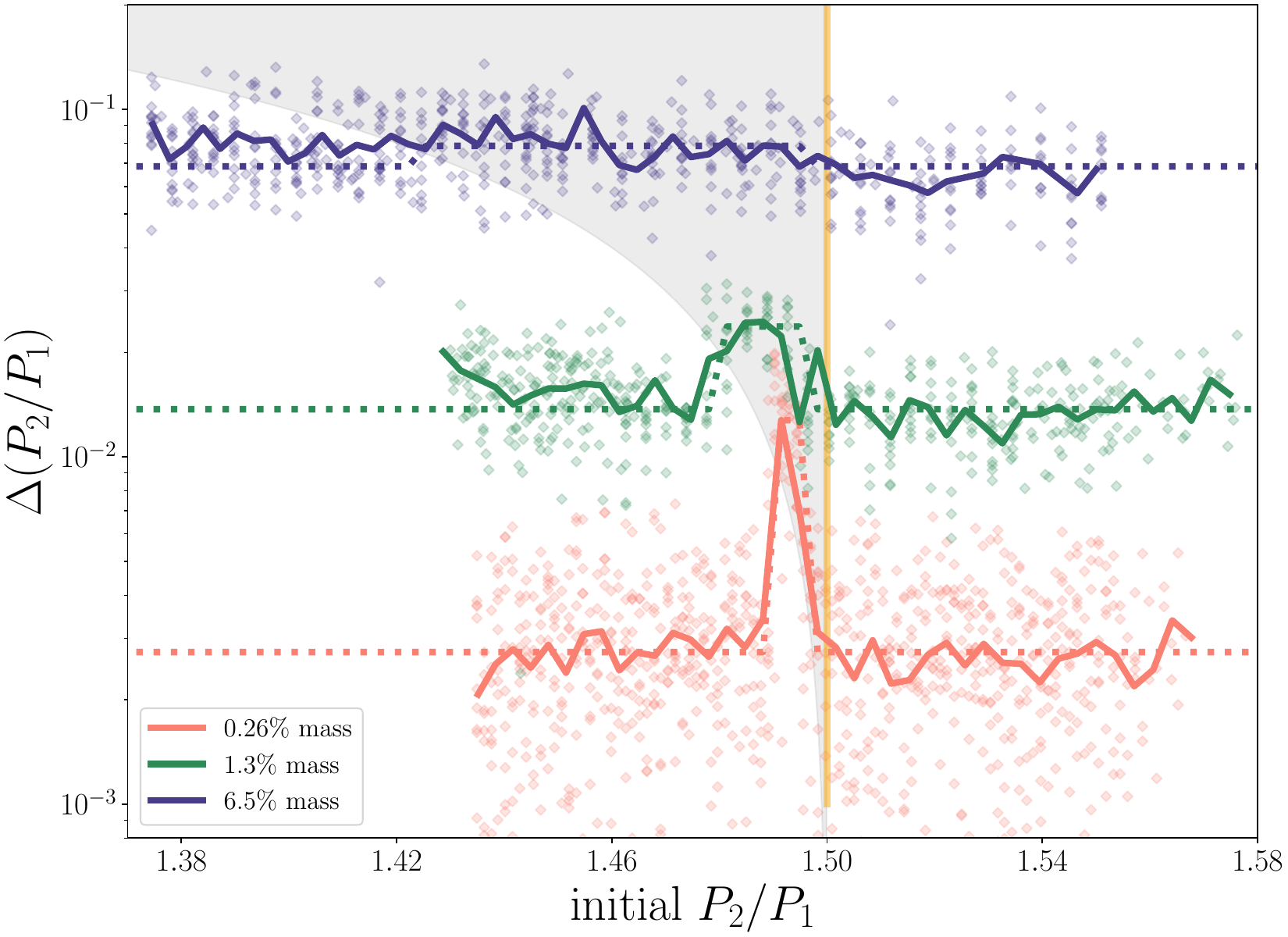}
    \caption{Repulsion of planet pairs, as a function of their initial period
      ratio (see Fig. \ref{fig:comparison} for the final one). Each group of colored points represent simulations with the same total mass ($\epsilon$) but a variety of ping-pong numbers ($N$). The thick colored lines depict the local mean, showing that the repulsion only depends on $\epsilon$.
      %All planet pairs experiencerepulsion ($\Delta (P_2/P_1) > 0$), the magnitudes of which   
     Points inside the shaded
      region encounter the 3:2MMR and therefore receive a resonance boost (eq. \ref{eq:dpp2}). Together, the repulsion is  
      well described by the dotted lines, which are the sum of 
eq.  \refnew{eq:dpp} and eq. \refnew{eq:dpp2}.   
      %      ping-pong mass is split into more bodies, the dispersion reduces roughly as a Poisson noise. 
    } 
    \label{fig:dp}
\end{figure}

\begin{figure*}
    \centering
\includegraphics[width=0.99\textwidth]{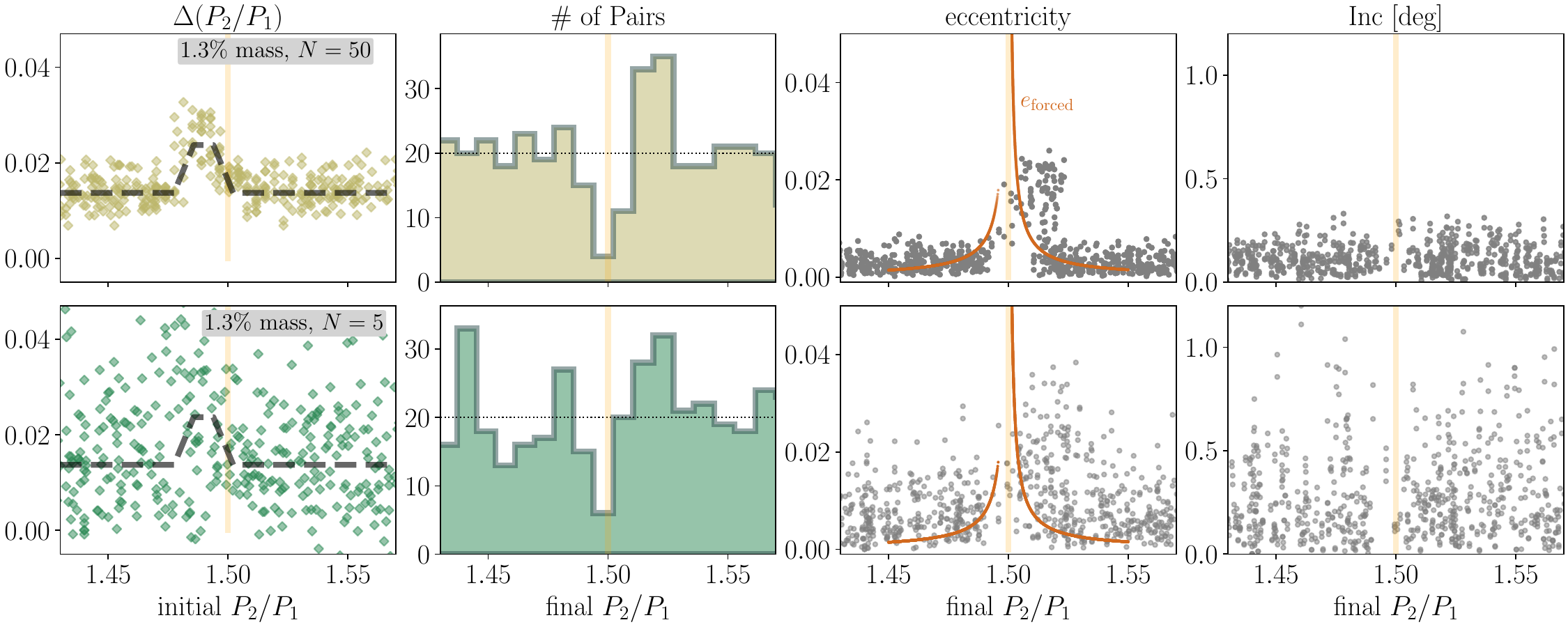}
 \caption{Comparing detailed simulation results
using light (top) and heavy (bottom) ping-pongs,  for the same total mass of $\epsilon=1.3\%$. Each ping-pong in the top panel weighs $0.08$ that of Mercury, while those in the bottom weighs $0.8$ Mercury. The left-most panels show the amount of repulsion, with the thick dashed lines indicating the so-called 'continuum limit' (eq. \ref{eq:dpp}-\ref{eq:dpp2}). The left-middle panels show the initial (dotted lines) and final (colored histogram) distributions in $P_2/P_1$. The bottom group exhibits a larger poisson noise, translating into a more jagged period distribution, and a less well-defined resonance asymmetry. 
   The panels to the right show the final planet eccentricities and inclinations. Planets in the upper group are only weakly excited, with eccentricity kicks from resonance crossing clearly visible. In contrast, the bottom group are excited to values that, if described by Rayleigh distributions, have modes of $\sigma_e \sim 0.008$ and $\sigma_{\rm inc}\sim \sigma_e/2 \sim  0.004$ radian (or $0.25\deg$). 
%   also   $i \sim e/2$.  
   %Inclinations, when fitted by Rayleigh distributioins, have modes of $0.25$ and $0.06\deg$, respectively.
   }
    \label{fig:comparison}
\end{figure*}

\begin{figure}
    \centering
\includegraphics[width=0.47\textwidth]{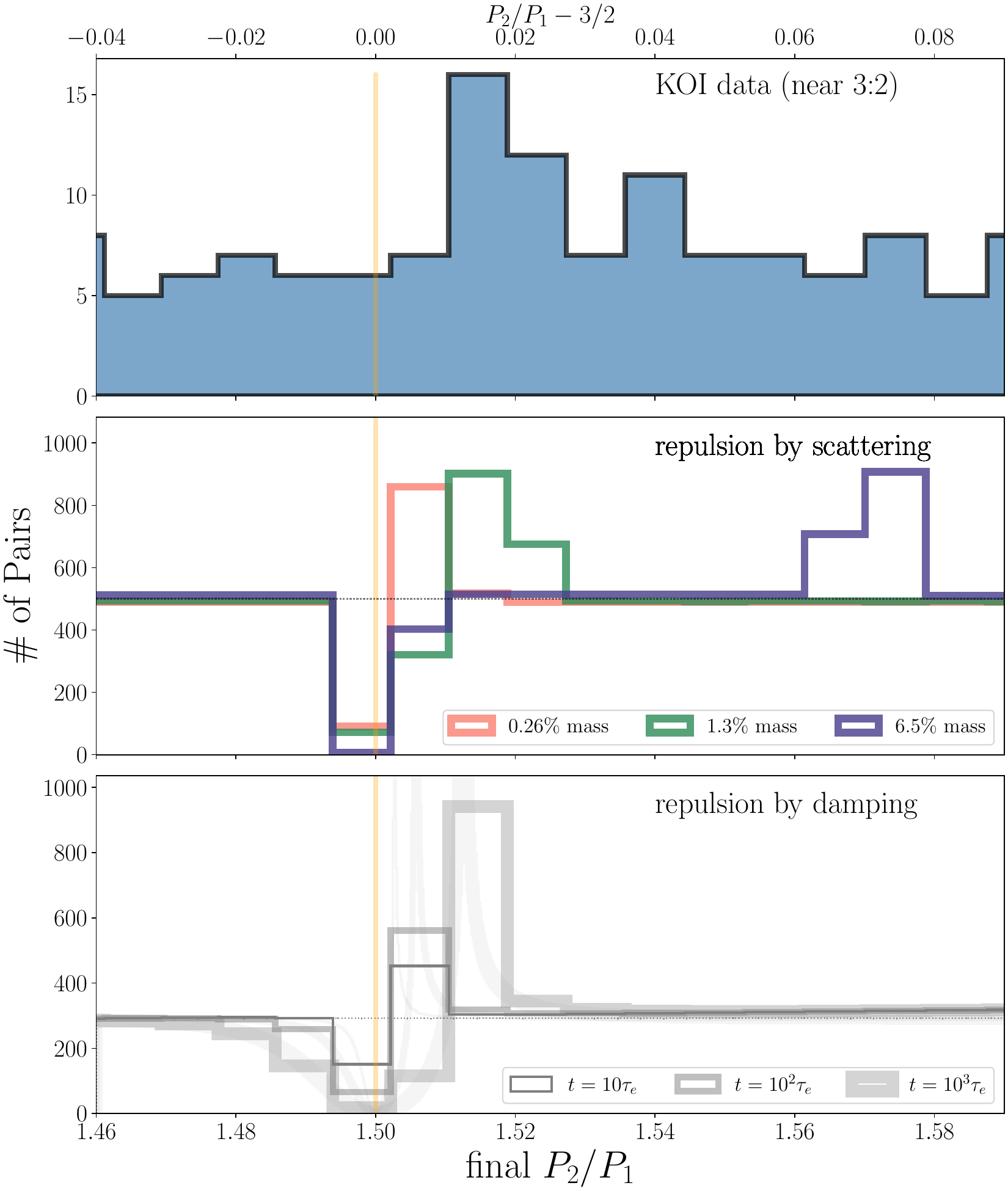}
    \caption{Resonance asymmetry near the 3:2MMR, observed (top panel) versus predicted (two lower ones, with the thin dotted lines indicate the initial distribution of period ratio).
    The middle panel shows the results of repulsion by scattering, for different values of $\epsilon$ (fractional mass in the planetesimals) and evaluated in the so-called 'continuum limit' (eq. \ref{eq:dpp}-\ref{eq:dpp2}).
%    \color{cyan}
%   Does  `continuum limit' mean that you are using analytic formulae, and then histogramming them? If yes, why is this called `simulated'?  It's also never stated  explicitly how exactly you use Eq 7.  Are you assuming that the resonance is centered on $P_2/P_1=1.5$? This could be done when discussing fig 8. (Note also that the dotted lines in Fig 8 are a little hard to see. The Eq 7 part looks trapezoidally-shaped, rather than rectangular, which I didn't understand. ). 
%    \color{black}
    The lower panel depicts repulsion by e-damping \citep{lithwickwu}, for different damping durations (measured in e-damping times), and in both histogrammed and unhistogrammed forms.
    \color{black}
  } 
    \label{fig:dp2}
\end{figure}

\section{Results}
\label{sec:requirements}

 %Our goal in this work is using ping-pong scattering to reproduce the observed pair statistics near MMRs, including the asymmetry and the e/inc distributions. 
 Here, we will simulate a large number of systems, varying parameters like ping-pong masses and numbers, and compare the results against observed pairs. For brevity, we focus on planet pairs near the 3:2 MMR.

\subsection{Simulations}

We continue to adopt the same planet parameters as before: $M_{\rm p1}=M_{\rm p2} = 8 M_\oplus$; planet radii $R_p = 1.5 R_\oplus$;
%$= 9600\km$; 
initially circular and co-planar orbits with the inner one at $a_{p1} = 0.1$au. We place the outer planet near the 3:2MMR, with an instantaneous period ratio
that is flatly distributed from $1.40$ to $1.60$. The word 'instantaneous' is only relevant within the libration width of an MMR. %We discuss this assumption in \S \ref{sec:discussion}.
%{\y did we really?}

The planetesimals (ping-pongs) are placed randomly from $a_{p1}/1.2$ to $1.2 \times a_{p2}$. Bodies in this range are found by \citet{chatter2022} to be actively interacting with the planets (their Fig. A1). The  initial eccentricities and inclinations of these bodies are assumed to follow Rayleigh distributions with modes $0.05$ and $0.05/2$, respectively.\footnote{Dynamically cooler ping-pongs are harder to excite to planet crossing orbits. But it may only be  a matter of time in a tightly packed planetary system.} We choose a total mass for the ping-pongs (mass ratio $\epsilon \equiv m/(M_{p1}+M_{p2})$, with $\epsilon$ ranging from $0.26\%$ to $6.5\%$) and spread the mass among $N$ particles (ranging from $5$ to $200$).
We expect the total mass to be the most important variable, with the value of $N$ only affecting the dispersion.
%We explore $M_p$ from 
%$20$ to $500M_\pluto$, 
%$0.4$ to $10 M_\male$,
%We choose $\epsilon = m/(M_{p1}+M_{p2})$ to range from $0.26\%$ to $6.5\%$ (so from $1$ to $20$ Mercury masses). 
%Our typical values for $N$ range from $5$ to $200$. 
%As a result, our ping-pongs have individual mass of order a Mars mass ($M_{\male} = 6.4\times 10^{26}\g$).
% $M_\pluto = 1.31\times 10^{25}\g$. so 50x smaller
We ignore mutual collisions between the ping-pongs by setting their physical sizes to zero. Moreover, the ping-pongs are taken to be pseudo-test particles, namely, they exert force on the planets but not amongst themselves. We integrate the system  until all ping-pongs have been accreted, or when we reach $10^6$ yrs.
%The latter is sufficient given that the collision time is of order $\sim 10^4$ yrs.

Fig. \ref{fig:dp} shows the collected results. All planet pairs experience repulsion, with a magnitude of $\sim \epsilon$ (eq. \ref{eq:dpp}). Pairs that encounter a first-order MMR during the process experience an additional boost that is of order the resonance width (eq. \ref{eq:delta}). As  $\epsilon$ increases, more pairs to the narrow of  MMR can encounter the MMR and enjoy this boost.
%As Fig. \ref{fig:dp} shows, the peak-trough asymmetry is the most clear when $\epsilon \sim 1\%$.

% Eq. \ref{eq:dpp} can be regarded as the repulsion in the continuum limit ($N \rightarrow \infty$).

To further illustrate some details, Fig. \ref{fig:comparison} 
compares two simulations of the same total mass ($\epsilon$) but different $N$. The run with a smaller $N$ (more massive ping-pongs) exhibit more scatter, translating into a  more subdued resonance asymmetry and an overall more jagged period distribution. The effects of $N$ are also seen in the final values for planet eccentricity and inclination. These quantities diffuse with time to near  equipartition values ($\sim \sqrt{m/M_p} e$), after a time comparable to the merger time. 
The simulation with 
heavier ping-pongs (Mercury-mass) produces an eccentricity excitation that can be described by a Rayleigh distribution with a mode of $\sigma_e = 0.008$, while the lighter group produces one that is  $\sim 3$ times smaller. 
In addition, the planets receive eccentricity kicks when they cross a resonance (see Appendix).

\subsection{Explaining the Data}

\begin{figure*}
\centering
\includegraphics[width=0.85\textwidth]{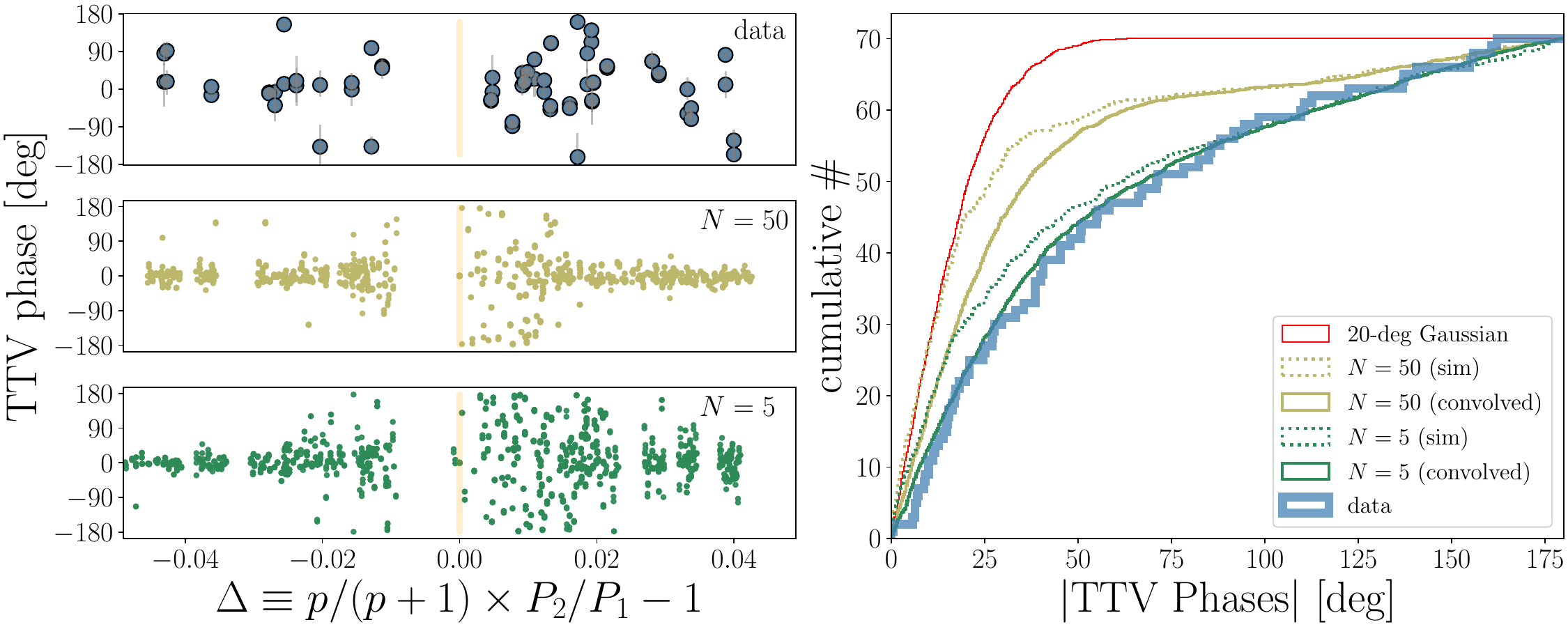}
\caption{Comparing TTV phases between data and simulations. For the observed values, we take the same 35 pairs from Fig. \ref{fig:freee} and express their period ratios as 
$\Delta = p/(p+1)\, P_2/P_1-1$ to capture their respective MMRs. The  simulations are the two sets from Fig. \ref{fig:comparison}. 
%  including all TTVs measured from different first order MMRs
%  \citep[data taken from Table 1 in][and for 3:2 and 2:1 MMR, removing pairs with phase uncertainties greater than $60\deg$]{Hadden2014}.{\w which paper?} Average $1-\sigma$ error is $\sim 20\deg$. 
%[The paper is HL14.  So I removed reference to HL17. Also, error bars are at 68\% confidence. ]
The left panels compare how the TTV phase depend on the distance to the MMR, and the right compares the cumulative distributions of the TTV phases. Here,
the red curve describes a population with zero free-eccentricities (convolved with the typical noise, a Gaussian of $20\deg$), and represents the prediction from repulsion by e-damping. 
The simulation with heavy ping-pongs, after convolving with the same Gaussian, agrees well with the observed TTV phase distributions (p-value $0.9$). The lighter ping-pong model, on the other hand, has a p-value of $0.0015$. This suggests that individual planetesimals should have $\sim$ Mercury mass.
}
\label{fig:ttvphase}
  \end{figure*}

%We will look at two observed aspects, the resonance asymmetry, and the free eccentricities and inclinations.
%\color{cyan} [this 1st sentence could be removed, given that discussion of first point was moved]  \color{black}

Fig. \ref{fig:dp2} shows the observed and the computed asymmetries around the 3:2MMR. The excess to the wide of the resonance is situated around $P_2/P_1 \sim 1.52$, and is best reproduced with a total ping-pong mass of $\epsilon \sim 1\%$.   A similar conclusion is reached to explain the pairs near the 2:1 MMR. In the same figure, we also contrast the scattering scenario against the damping scenario. Using the formulation from \citet{lithwickwu}, one finds that an extensive period of eccentricity damping, of order $10^3$ e-damping times, is required to reproduce the same asymmetry (see \S \ref{sec:intro}).

Observationally, both the 3:2 and 2:1MMRs show an excess of pairs to the wide of the resonances, but only the 2:1 shows a clear deficit to the narrow (Fig. \ref{fig:dp1}).  %because any pairs crossing this region should be carried across quickly by resonant dynamics. While this 
We will discuss this issue in \S \ref{subsec:caveat}.
%{\y moved the messiness to \S \ref{subsec:caveat}}

Now we turn to consider planet eccentricities. 
%In our model, the planets are initiated with zero eccentricities. These grow to equipartition values ($\sim \sqrt{m/M_p} e$) after a time comparable to the merger time.  In Fig. \ref{fig:comparison}, the simulation with  the heavier ping-pongs (Mercury-mass) produces an eccentricity excitation that can be described by a Rayleigh distribution with a mode of $\sigma_e = 0.008$ (and $\sim 3$ times smaller for the lighter group). 
\citet{Hadden2014} reported, for a sample of 54 planet pairs, 
%that derived from a sample of 44 TTV planets by \citet{wulithwick}, 
a Rayleigh mode of $\sigma_e = 0.018$. This is a couple times larger than that in our simulations with Mercury-mass ping-pongs. These eccentricities are derived from the observed TTV phases (see \S \ref{sec:intro}), so a more direct comparison should be done in terms of TTV phases. The data in Fig. \ref{fig:freee} consists of 35 pairs, which are the better measured subset out of the 54 pairs in \citet{Hadden2014}.
We compare these against those from our simulated pairs from Fig. \ref{fig:comparison}. For the latter, we extract TTV information 
%from the two groups of simulations in Fig. \ref{fig:comparison} 
as follows. For each planet, we express its transit times as the sum of four terms: a linear trend, an offset, and two sinusoidal terms that account for the amplitude and phase of the TTV \citep[Eq. 18 in][]{lxw}. 
The four coefficients of these terms
are extracted via linear least-squares 
\citep{1986nras.book.....P}. We then convolve the simulated TTV phases with a measurement error of $20\deg$ and compare the distributions in 
Fig. \ref{fig:ttvphase}. We find that only the simulations with heavier ping-pongs (Mercury-mass) can explain the observed spread in the TTV phases.
%We find that run with more massive ping-pongs (individual mass $\sim 0.8$ Mercury masses) provide a good fit to the observed distribution, especially when the simulation data is convolved with the error (assumed to be $20\deg$ Gaussian).
%In contrast, repulsion by e-damping  predicts no spread in the TTV phases and is consistent with data. 

The heavy ping-pong run also yields  planet mutual inclinations that are nearly Rayleigh with a mode of $0.25\deg$. This value is a few times smaller than the $0.8\deg$ inferred for the high-multiple systems \citep{zhuwu}, and a larger factor away from the $1-2\deg$ inferred for general pairs \citep{fabrycky}. We  consider this a partial success of our theory:  the ping-pongs can impart some non-coplanarities to the planets, and it is possible that further dynamical instabilities, not captured in our theory, have set in to increase the dispersion.
% he used all pairs
%\x{So we fail in this respect.} \x{It is possible that mutual inclinations can also be excited by stellar obliquity \citep{spalding16} and/or inclined perturbers \citep{pulai}.}
%This leads one to suspect an earlier episode of excitation with more massive ping-pongs.

In summary, our simulations suggest that a handful of Mercury-mass ping-pongs can explain the resonance asymmetry, the magnitudes of the free eccentricities, and partially, 
the mutual inclinations.

\section{Discussion}
\label{sec:discussion}

We propose that ping-pong repulsion is responsible for the observed MMR asymmetry.  Repulsion should have occurred across all pairs, not
just those close to an MMR. But its effect is clearly  discernible only near
an MMR.
%since this process should not occur only near some special period ratios, and since the observed asymmetry is of order unity, we suggest that ping-pong repulsion must have occurred universally across all pairs. 
%Its effect is hard to discern except when close to an MMR.

In the following, we provide an astrophysical context that may give rise to such a universal behaviour (\S \ref{subsec:story}). We will also consider how the dynamics changes when the planets have gaseous envelopes (\S \ref{subsec:gas}). In \S \ref{subsec:messy}, we point out that accretion is likely complicated.
We discuss previous works in \S \ref{subsec:chatterjee}, and an important caveat in  \S \ref{subsec:caveat}.

\subsection{A plausible story}
\label{subsec:story}

Our calculations suggest the presence of a handful of Mercury-massed objects (radius $R \sim 2300\km$ if at Earth composition)
that make up $\sim 1\%$ of the planet masses, at late stages of planet formation. What astrophysical context can naturally give rise to this population of bodies?

The following estimate is thought-provoking. We have ignored ping-pong mutual collisions in previous sections. But they do collide.
%For a total mass ratio $\epsilon$ and a ping-pong radius $R$, 
Orbit-crossing ping-pongs should collide with each other at  a timescale 
\begin{equation}
        t_{\rm coll}
        %\sim {1\over {2N}} P_{\rm orb} \times {{2 \pi a^2 }\over{\pi   R^2}} 
        \sim 3\times 10^5  {\rm yrs}\,
    \left({{a}\over{0.1 {\rm au}}}\right)^{3.5}\, 
    \left({{\epsilon}\over{1\%}}\right)^{-1}\, 
    \left({{R}\over{2300 {\rm km}}}\right)
   \, .
    \label{eq:tcoll}
\end{equation}
Within uncertainties, this is comparable to the planet accretion time of ping-pongs, $t_{\rm merger} \sim 4\times 10^4$ yrs (eq. \ref{eq:tmerger}).
Is this a coincidence?

Let us first imagine a situation where $t_{\rm coll} \ll t_{\rm merger}$, i.e., ping-pongs are small and collide frequently with each other. This may introduce collisional cooling to their orbits, grind them down to smaller debris, and in general forestall orbit-crossings with the planets.
%if the bodies are pulverized into smaller debris which collide even more frequently. As a result, can forestall planet scattering. 
Only when the ping-pongs grow large enough so that $t_{\rm coll} \sim t_{\rm merger}$, can the scattering/accretion proceed.

Let us now consider the opposite limit when the mass is dominated by too few large ping-pongs and $t_{\rm coll} \gg t_{\rm merger}$. These will be quickly accreted but the accretion process is messy (see \S \ref{subsec:caveat}): grazing impacts and tidal disruptions may allow some of the ping-pong mass to escape in the form of small fragments. These fragments may be in the collisional regime ($t_{\rm coll} \ll t_{\rm merger}$) and the previous discussion applies. 

As a result, one can imagine that, when $\epsilon$ drops from unity to zero over many decades, the dominant ping-pongs continuously evolve in size to maintain $t_{\rm coll} \sim t_{\rm merger}$, or a ping-pong size $R \propto \epsilon$,  and a ping-pong number, $N \propto R^{-2} \propto \epsilon^{-2}$.  For our parameters, $N \sim 5 \times (\epsilon/1\%)^{-2}$.

In such a picture,  $\epsilon \sim 1\%$ is special. First, this makes the resultant MMR asymmetry clearly visible  -- a smaller $\epsilon$ 
would produce too weak a 
%the repulsion is too weak to leave much of a signature. The number of pairs that will be pushed towards and can jump across the MMR is small 
repulsion and very few pairs can be pushed towards and jump across the resonance (see the case with the lowest $\epsilon$ in Fig. \ref{fig:dp2}). 
Second, the asymmetry signature is less stochastic at this value of $\epsilon$ because $N \sim 5 \gg 1$ -- at a much larger value of $\epsilon$, ping-pongs will be fewer in number  ($N \ll 5$) and almost planet-like in size. They would cause large repulsion but the signature is too stochastic to be recognizable (compare the two rows in Fig. 
\ref{fig:comparison}). In fact, ping-pong repulsion in this stage may give rise to a nearly flat period distribution, as observed in generic planet pairs; it can also leave a finger-print in the eccentricity/inclination excitations.  

So in conclusion, we suggest that the late stage of planet formation is a time of clearing of a substantial amount of debris. The ping-pong generation with $\epsilon \sim 1\%$ leaves the last and the clearest MMR signature, explaining naturally the observed asymmetry at $1-2\%$ away from exact MMR. 
More massive early generations may instead account for the nearly feature-less pair spacing in {\it Kepler} planets.

\subsection{Gas: accretion and erosion}
\label{subsec:gas}

Many of the {\it Kepler} planets possess low-mass hydrogen envelopes (mass of order $1\%$), the so-called mini-Neptunes. It is likely that they accrete these envelopes before the gas disks disperse.

In the following, we consider how the presence of a gaseous envelope affects ping-pong accretion, and, importantly, whether the ping-pongs can strip the planets of their envelopes. This discussion assumes that the scattering occurs after gas accretion. This is possible, despite the fact that the
disk lifetime (Myrs) is much longer than the time to clear the ping-pongs (eq. \ref{eq:tmerger}).
%If gas accretion preceeds ping-pong accretion, as long as gas does not affect the orbits, all our previous results stand. 
%On the other hand, it is 
For instance, dynamical friction (either from gas or from mutual collisions) may place the ping-pongs into psuedo-stable orbits (like the gray crosses in Fig. \ref{fig:dp0}), so they may reside there for millions of years before being dislodged by weak perturbations. 
%\footnote{One way to place ping-pongs in these psudo-stable orbits is mutual collisions. They reduce orbital excitations and settle the bodies into the so-called 'minimum energy' state.} They remain inert (avoid orbit-crossing) for at least 1 Myrs. 
%if former, formation inside out, the original near resonant pair may be compressed, trapped into MMR, not seen; 

%\footnote{This is in contrast to the solar system, where the clearing stage lasts of order $10^8$ yrs, both in the terrestrial zone and in the Neptune region.}

%\x{Alternatively, there could be ping-pongs that are sent inward by the outer system, when it undergoes dynamical clearing. But it is unclear whether these repel the planet pair.}
%if so, does re-organization of the planetary system starts outside in?
%could even be drifted in by Yarkovsky effect?

\subsubsection{Gas Envelope and Accretion}
\label{subsec:accreter}

Previously, we have set the planet's accretion radius to be its core size. Now we consider mini-Neptunes. The observed hydrogen atmospheres contain about a percent of the planets' masses, but expand their physical sizes by a factor of 2 .
%($R_p \approx 2 R_{\rm core} = 3 R_\oplus $). 
%(so $2^3-1^3=7$ times the volumn) this means 
The mean density of the atmosphere is then smaller than the solid (core) density, by a factor of  $\rho_{\rm atm}/\rho_{\rm solid}\sim 10^{-3}$. A solid planetesimal travelling through the atmosphere will lose a fraction of its energy that is roughly the ratio of the encountered mass to its own mass,
\begin{equation}
    {{\delta E}\over E}\sim {{\rho_{\rm atm}}\over{\rho_{\rm solid}}} \times \left({{R_p}\over{R}}\right) \sim  \left({{25\km}\over{R}}\right)\, ,
\end{equation}
with merger occurring whenever $\delta E/E \sim 1$. 
So for our preferred ping-pongs (Mercury-mass, $R \sim 2300\km$), the atmosphere presents little ability to capture. This justifies our using the super-earth size for merger.

\subsubsection{Survival of the Gas Envelope}
\label{subsec:survival}

Would the ping-pongs remove the gas envelopes? This is a relevant point because the ping-pong mass fraction ($\epsilon \sim 1\%$) is comparable to the envelope mass fraction of the mini-Neptunes. 
%So it is relevant to consider the efficiency of atmospheric removal (mass removed vs. mass accreted) for a given impactor.

Let us first consider a grazing impact where the ping-pong misses the solid core. The encounter speed is very supersonic, so the ping-pong simply punches a hole in the atmosphere, carrying away with it the 
fraction of the envelope that is in its path, or of order $(R/R_p)^2 \sim 1\%$ of the envelope mass, with $R_p$ taken to be $4 R_\oplus$. Multiple impacts may enhance this fraction, but the ping-pong will likely experience an impact that hits the solid core and be absorbed, before it is able to strip off the envelope.

For ping-pongs that hit the solid core,  
%The column density of the atmosphere is $\sim M_{\rm atm}/(4\pi R_p^2) \sim 2\times 10^7 \g/\cm^2$.  Assuming a rocky composition (bulk density $\rho_{pp} \sim 2.7 \g/\cm^3$), any bodies larger than $\sim 30\km$ can hit the ground. 
a shock emerges at the antipole of the impact point, accelerating the gas there to escape. According to 
\citet{almog1} and their corrected results \citep{almog2}, even for an impact speed that is twice the surface escape ($\sim 25$km/s for our super-Earths) %orbital $100$km/s, take $e=0.5$, so $v_{\rm impact} \sim 2 v_{\rm esc}$).
and an impactor that is $\sim 1/4$ in radius to the target, 
%when the impactor is smaller by $10$ than the target (Pluto radius is $0.1\times$ super-earth radius), it is able to remove $\sim 10^{-6}$ of the atmosphere (assume to be $1\%$), or $\sim 10^{-5}$ of the impactor's mass. 
%This inefficiency is related to the small ground movements induced by a small impactor.  {\w but \citet{schilichting2015} will claim opposite?}
only about $10^{-3}$ of the envelope can be removed. This is inmaterial.

%To make the impactors as damaging as possible, we put all mass ($\sim 1\%$ planet mass, see above text) into a single body, leading to $R \sim 0.2 R_p$. At $v_{\rm impact} = 2 v_{\rm esc}$, this body can remove $\sim 10^{-3}$ of the atmosphere, or $10^{-3}$ of its own mass.

%Two other comments. If we instead assume that the impactors are smaller than $30\km$, they burst in mid-air. \citet{burst} argued that again, the efficiency of removal is at best a few percent. \citet{heatloss} considered the heat deposition by impactor in removing the atmosphere and again it is inefficient (Fig. 5 of that paper).

\subsection{Messiness of the Accretion Process}
\label{subsec:messy}

The process of merger with a solid core, on the other hand, can be a messy business.

For one, tidal disruption, as opposed to a complete accretion, may interfere. While the ping-pong density can be terrestrial (we take $\rho_p = 5\g/\cm^3$), that of the super-Earth is higher due to gravity compression, and can be $\rho_{\rm core} = 12\g/\cm^3$ (corresponding to $8M_\oplus$ and $1.5 R_\oplus$). This enhances the cross-section for complete tidal disruption over that for accretion. For instance, for
a high velocity impact of $v_\infty = 0.5 v_{\rm esc}$ (this corresponds to $e\sim 0.15$), 
%watanabe92: factor 1.7 is down to 1 when initial velocity is $0.5 v_esc$. 
\citet{boss1991,watanabe92} predict a destruction distance of,
\begin{eqnarray}
R_{\rm Roche} & \sim &  1.0 R_{\rm core} \times \left({{\rho_{\rm core}}\over{\rho_{p}}}\right)^{1/3} \nonumber \\
& \sim & 1.3 R_{\rm core} \left({{\rho_{\rm p}}\over{5\g/\cm^3}}\right)^{-1/3} \left({{\rho_{\rm core}}\over{12\g/\cm^3}}\right)^{1/3}
\, ,
\end{eqnarray} 
This distance  expands further by a factor of $1.7$ if the encounter is parabolic  \citep{sridhartremaine,watanabe92}.
%, and to $2.5$ for a circular orbit \citep{Chandrasekhar,..??}.
Tidal disruption can return a fair fraction of the ping-pong mass to stello-centric orbits. These are not directly accreted by the planet.

But even if the ping-pong manages a direct hit on the core, all is not clear.
Simulations by \citet{AgnorAsphaug,Genda2012} and others have shown that, for $v_\infty \geq 0.5 v_{\rm esc}$,
% I used Fig. 8 of Genda2012, 1.2 v_esc for small mass ratio(horizontal axis = 1), at 45deg
% though average should be more like 60deg; 
%%apparently everyone in that business use impact velocity, not velocity at infinity, so $\sqrt(1.2^2-1) \sim 0.67$
the more likely outcome of a collision is "hit-and-run", where, instead of perfect accretion, the ping-pong is only partially absorbed, with much of the rest escaping as a stream of debris. 

%$v_{\rm esc} \sim 26\km/\s$

%As an illustration,  we will adopt $\rho_{pp} = \rho_\oplus$, where $\rho_\oplus = 5.5\g/\cm^3$ is the density of the Earth; and we take $\rho_P \sim \rho_\oplus (M_p/M_\oplus)^{1/4}$  \citep{density??}. This density contrast dictates that tidal disruption is a more common occurrence than direct impact. 

Combining these two aspects, we have the following picture for collisions in a gas-free environment.  At low speeds ($v_\infty  \leq 0.5 v_{\rm esc}$), the ping-pongs are predominately tidally destroyed. At higher speeds, tidal disruption is less active but ping-pong accretions are largely hit-and-run. Accretion is messy and complete clearing may take longer than our simple estimate (eq. \ref{eq:tmerger}). How this affects the dynamics requires a closer look. 
%One, the effective accretion time is prolonged, perhaps by a factor of few. Second, much of the direct impact mass may be in the form of small bodies, even if the original ping-pong is much more massive.  {\w some of the debris may be ground down to micron-sized particles and blown out by radiation pressure? and/or providing e-damping?}

\subsection{Previous Works}
\label{subsec:chatterjee}

It was first proposed by \citet{chatter2015} that scattering with planetesimals can repel planet pairs. They were interested in studying whether planet pairs initially trapped in MMRs can break free by scattering with the planetesimals. They reported pair repulsion -- the root cause of which we 
 elucidate here -- and that the amount of repulsion scales with the mass of planetesimals.
%(much of the wide belt is not stirred up to orbit corssing), 
%fig. 7, take alpha=0
%However, this work did not focus on understanding the fine features near the resonance and detailed comparison with data, and the large belt mass they advocate casts a pessimistic light on the mecha

This problem is further studied by \citet{chatter2022}, with the specific aim of reproducing the resonance asymmetry. Like in the current study, they initialized the period ratio of the planet pairs as a flat distribution (i.e., not only in MMRs). Within their wide belt of planetesimals, only about $1/5$ are destabilized by the planet into orbit crossing (their Fig. 2). Accounting for this inefficiency, their $m_d/m_p = 0.1$ model has an effective $\epsilon = 2\%$. For such a value, we expect an orbital repulsion of $2.2\%$ (eq. \ref{eq:dpp}), while \citet{chatter2022} only found $0.3\%$ (their Fig. A1), a repulsion efficiency that is almost one order of magnitude lower.
% I got this by taking their 0.5 disk, epsilon goes from 0.115 to 0.130, so divide by 5 to get $0.3\%$
Moreover, their simulations fail to reproduce a prominent resonance asymmetry (their Fig. 7 for 3:2 and Fig. 8 for 2:1), in contrast to our results here (Fig.  \ref{fig:comparison}).
They do find that disks with even higher masses could produce some asymmetries, but at locations differing from that is observed.

There may be a number of factors that could help explain our differences.  For instance, their simulations are expensive because they use a large number of planetesimals, 
so they can only perform a small number of runs.  This may introduce statistical fluctuations and obscure the asymmetry. Also, their planet sizes (based on an ad-hoc mass-radius relation) are likely larger than ours (only the core size), leading to rapid accretion. Lastly, the massive belt of ping-pongs adopted in \citet{chatter2022} may exert dynamical friction on the planets,
obscuring the effect of ping-pong repulsion.

%\citet{chatter2022} choose to investigate the limit of small ping-pong masses. The typically use $5000$ bodies to represent disks that is, say, half of the planet mass. This corresponds, to our case ($1.3\%$ planet mass) of $N \sim 100 $, even lower than the lower panel in Fig. \ref{fig:comparison}. The low ping-pong mass brings about low values of planet eccentricities and inclinations at equip-partition, precluding them from explaining the TTV phases (Fig. \ref{fig:comparison}). 

%Unlike our simulations, \citet{chatter2022} adopted  very small planetesimal masses, and are effectively in what we call the 'continuum limit'. In this limit, Fig. \ref{fig:dp2} shows that a mere $\epsilon=1.3\%$ can reproduce the observed asymmetry.
% This mass is represented by $\sim 1000$ bodies, so their large simulations are effectively in what we call the 'continuum limit'. 

Due to the failure of \citet{chatter2022} to produce the MMR asymmetry, they argued that there must be a significant population of primordial resonant pairs (though only in 3:2, not in 2:1). This population does not seem warranted in our study to reproduce the asymmetry. However, there is a related conundrum on exact-resonant pairs, we address this point below.

\subsection{Caveat} %Flanking bodies: planets and planetesimals }
\label{subsec:caveat}

We note that currently there are a number of exoplanet pairs in libration in the 3:2 and 2:1 MMRs, mostly in systems with resonant chains \citep[see, e.g.][]{daifei2023}.
%%Many other planets with similarly low Δ also have a resonant chain of planets.
If these pairs are removed from the statistics, the region to the narrow of 3:2MMR will exhibit the expected deficit.\footnote{These resonant pairs are also absent from the low-noise TTV sample in Fig. \ref{fig:ttvphase},
as their TTV signals are 'noisy' due to the presence of a third body.}
%\color{black}
%An interesting feature in Fig. \ref{fig:comparison2} is that the 35 good-quality TTV pairs seem to avoid the resonance, just like our simulated pairs do.  is interesting to note that these 35 pairs clearly avoid  \color{black} {\y checked, indeed the more-exact resonance pairs are excluded due to large error-bars}
%Moreover, Fig. \ref{fig:dp2} shows that pairs that have crossed the MMR are more eccentric. Current data seem consistent with this but the errors are large.}

These resonant pairs appear to be results of convergent migration, and, for some reason, have avoided the evolution outlined in this paper. The fact that  they exist mainly in resonant chains may provide a clue, but one that has yet to be deciphered.

In this work, we have only considered ping-pong repulsion of an isolated planet pair. In reality, pairs are embedded in a ladder that are all jostling to repel each other. It is unclear what the outcome of this competition is. 
%What is the impact of planets besides the pair? If the ping-pongs are stored in all orbital ranges, they always lead to local pair repulsion. Then flanking planets causes competition in orbital expansion. This could reduce the efficiency of repulsion by order unity. Another effect is the absorption of ping-pongs when they orbit-cross with flanking planets. 

%Planetesimals flanking the pair is another complication. As shown by Fig. A1 of \citet{chatter2022}, only those within our calculated range can be activated by the pair and interact with them. However, flanking planets can again catapult remote ping-pongs to their domain, leading to unclear effects. These should be investigated.

%\begin{enumerate}
%\item initial planets have circular orbits; circular is not too important... flat is also not too important; 
%\item initial planetesimals nearly circular, coplanar orbits, also not too important
%\end{enumerate}

\section{Conclusion}
\label{sec:implications}

The process that produces the observed asymmetry around MMRs may be the last imprint of planet formation. As so, it is an important key  to unlock mysteries of planet formation.

We consider the situation where, after the {\it Kepler} planets were fully formed, and after the disappearance of the gaseous disks, there may still remain some  planetesimals. These are then gradually cleared away by a combination of planet scatterings and accretion. We show that this naturally repels a planet pair, by of order the mass fraction in the small bodies. We elucidate the origin of this repulsion: due to the larger phase space at high eccentricities and inclinations, the small bodies tend to diffuse to regions of larger angular momentum deficit. In other words, they absorb, fractionally, more energy than angular momentum from the planets. 
An originally circular planet pair can only adjust by pulling apart, with the outer body receiving a donation of energy and angular momentum from the inner one. The small body acts as the agent for this transaction. This dynamics is analogous to the spreading of an accretion disk, when its orbital energy is dissipated into heat.
Given the prominence of the MMR feature, we expect 
this process to have occurred universally among all planet pairs, and to only become recognizable when the pair receives a MMR boost. Moreover, a planetesimal belt of order $1\%$ of the planet mass is required to reproduce the magnitude of the resonance asymmetry.

Unlike repulsion by eccentricity-damping ('resonant repulsion'), this mechanism can excite planet eccentricities and inclinations. To account for the values of eccentricity and mutual inclinations reported by transit observations, the planetesimals should be  of order Mercury in mass, with a total number that is in the single digits.
We outline an evolutionary pathway to achieve such a residual planetesimal population near the end of the planet formation era. 

We argue that accretion of these ping-pongs would not lead to appreciable atmospheric loss. The final accretion may be messy, involving either tidal disruptions or hit-and-run collisions.
%\x{How the debris from these processes affect the planet pair is unknown.} 
In addition, we have not yet explored how ping-pong repulsion alters the orbits of a trio (or more) of planets. 
%\x{We have not investigated the imprint of scattering on a chain of planets, but we do not expect, e.g., a planet triplet to lie along 3-body resonances.} 

%Their scattering with the planets then shift the pair spacing from the original feature-less shape to ones with an asymmetry near the MMRs, hence leaving us a clue to the genesis epoch. In the mean time, the planets' originally cold orbits are excited to have small free eccentricities, as is observed. 

If such a story holds, we suggest that the late
stage of planet formation is a time of clearing of a substantial amount of debris of lunar-to-Mercury mass planetesimals.
%In a way, these ping-pongs remind one of the spring argument in accretion disks.  just like the magnetic field in MRI, communicating between two parts of the disk by the spring leads to divergent spreading. the physics is different, here, ping-pongs statistically want to diffuse to higher-e phase space.

%Coupled with the rapid crossing at first-order mean-motion resonances, this can produce the observed asymmetry. The excitation by scattering also explains the presence of free inclinations and eccentricities, as is observed. 

\begin{acknowledgments} We are grateful to Hanno Rein for answering questions on REBOUND, and to Eugene Chiang, Sam Hadden, Scott Tremaine for discussions.  YW is funded by the NSERC discovery grant, 
RM is funded by NASA grants 80NSSC18K0397 and 80NSSC21K0593, and YL is funded by NASA grant 80NSSC23K1262. 
\end{acknowledgments}

\bibliographystyle{aasjournal}
\bibliography{scatter}{}

\appendix

\section{Kicks from a divergent encounter with a mean motion resonance}
\label{sec:app}
%-- analytic formulas}

\begin{figure}
    \centering
 \includegraphics[width=0.45\textwidth]{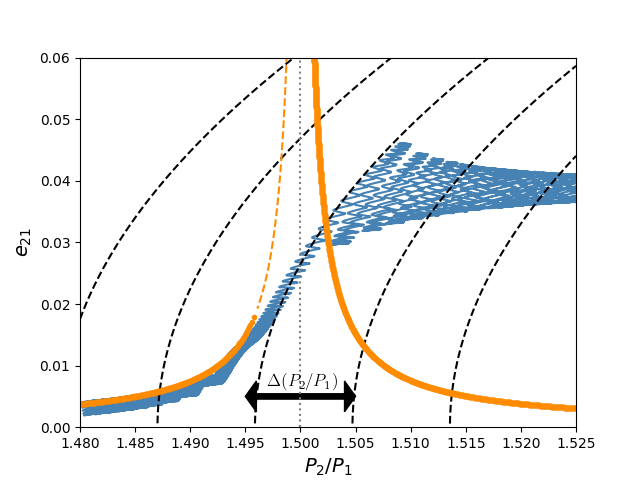}
    \caption{Repulsion associated with a MMR crossing is related to the unstable fixed-point. The horizontal axis shows $P_2/P_1$ and the vertical axis the relative eccentricity (a reduced variable, see text).  The blue curve shows the REBOUND integration of an equal-mass planet pair, being gradually pushed apart across the 3:2MMR. The black dashed curves correspond to constant Brouwer integrals, while the orange curves are the fixed points for eccentricity, with the solid orange curve representing the stable fixed-point and the dashed curve the unstable one.  
%    with simplified pushing: Two planets are forced to cross the 3:2 resonance, by applying an artificial $\dot{a}$ on the inner planet. The blue curve shows how    the pair's relative eccentricity ($e_{21}$)  evolves as the period ratio increases. 
As the pair gradually moves from left to right, they encounter the unstable fixed-point.  Within one libration time, the pair jump across the MMR. The associated kicks in period ratio and eccentricities are discussed in the text.
}
\label{fig:simplepush}
\end{figure}

When two planets migrate across an MMR divergently, they acquire  kicks to their eccentricities and  period ratio
\citep{Henrard:1983,murray}.
%\color{cyan} [Need refs. Renu: do you know an early ref for divergent kicks?
%{\renu Ans: \cite{Henrard:1983}, however, in common with most of the literature on resonant encounters, they did not explicitly discuss kicks to period ratio, which is implicit in their analysis}]
%\color{black}
We illustrate this with a REBOUND integration.
We gently push two planets apart, across the 3:2MMR, by imposing an  inwards $\dot a$ on the inner one. The result is the blue curve in Fig. \ref{fig:simplepush}, which shows the two planets'
$e_{21}$
(the magnitude of the difference of their eccentricity vectors)
versus their period ratio.
The masses of the planets are each 8$M_\oplus$, and
they are initialized with $P_2/P_1$ inward of the MMR.
The initial $e_{21}$ has zero free 
component---i.e., it is placed at its analytic fixed point, 
which is traced out by the orange curve.
%represented by the orange curve
% with
% zero free-eccentricity; i.e., we set $e_{21}$
 %{\renu [how was zero free eccentricity imposed?]}. They
 The planets
  gradually evolve to 
%$B$ is no longer constant. Instead, provided $\dot{a}$ is sufficiently 
%slow,
%then provided the forcing rate is sufficiently slow
 larger $P_2/P_1$.\footnote{Although the imposed $|\dot{a}|/a$ used for Fig. \ref{fig:simplepush} is small relative to the resonant libration frequency, it is much larger than realistic migration values.
%from ping-pong scattering (averaged over long timescales).
%{\renu [does this refer to the individual kicks of ping pongs or some smoothed/average? may be worth clarifying this point.]}. 
That artificial enhancement makes the resonant librations in
Fig. \ref{fig:simplepush} more visible, as it prevents successive librations from overlapping too much.}
The kicks in $e_{21}$ and $P_2/P_1$ upon resonant crossing are evident in the figure.

The behavior in Fig. \ref{fig:simplepush} can be quantified by comparing with what
  happens for non-migrating
planets. 
A planet pair near a MMR
may be reduced to 
 a Hamiltonian with two variables, 
 $e_{21}$ and $\phi$,  where
 $\phi$ is the resonant angle
\citep{1984CeMec..32..307S,1986CeMec..38..335H,deck13real}.
The MMR gives rise to  oscillations in both  $e_{21}$  and 
$P_2/P_1$, which are coupled via
a constant of the motion, $B$:
\be
 B\approx
 {P_2\over P_1} -{3p\over 2}\left({p+1\over p} \right)^{5/3}e_{21}^2\, ,
 \label{eq:brou}
 % \sigma^2 - {2\over 3k}{\Delta p\over p_{\rm res}^{5/3}} \ .
 %\left({{\cal P}\over {\cal P}_{\rm res}}-1\right) \ ,
 \ee
%kick in the period ratio is determined from the kick in $e_{21}$ by constancy 
%of the Brouwer integral, which near the $p:(p+1)$ MMR is
for a $p+1:p$ MMR. 
The expression in Eq. \ref{eq:brou} follows from the
conservation of
$\Gamma'$ in \cite{deck13}, as approximated
in their Eq. 3.  
Their approximation assumes that $p$ is large in the Laplace coefficients, but we find nonetheless that it is adequate for the 3:2 and the 2:1.  
See \cite{deck13real} for a more thorough discussion of the approximations involved. 
%\color{cyan} [Renu suggests I be more consistent in my approximations, and set $p+1\rightarrow p$ everywhere. But that makes the expressions less accurate for the 2:1 and 3:2. My feeling is that for our purposes we would prefer more accurate expressions, at the expense of cosistency.  Note also that Deck et al 13 show that the Hamiltonian is nearly independent of mass ratio, subject to assumptions that are moderately good. My inclination is to leave it like this.]\color{black}
The constant $B$ 
 is often referred  to as  the ``Brouwer integral''  \cite[e.g.,][]{2005AJ....130.2392H}. 

 Fig. \ref{fig:simplepush} shows  curves of constant $B$  near the 3:2 MMR.
The resonant oscillations are visible as the motion that nearly tracks the constant $B$ curves. 
 As the planets are pushed apart ($P_2/P_1$ increases), they adiabatically follow the fixed-point solution,
 determined
 using
the Hamiltonian given in  Eq. 1 of
\cite{deck13}
for a given value of $B$ (or $\Gamma'$).
%\label{fig:simplepush}
However, at a certain value of $P_2/P_1$ ($\sim 1.4955$  for our case), the fixed point becomes unstable. 
 The  pair jump within a single libration time to $P_2/P_1\sim 1.505$. 
% [That addendum is not really necessary. But I added it because the $\dot{a}$ in Fig 13 was made artificially big so that the figure was clearer, and with that big $\dot{a}$, it gives a slightly wrong $\Delta(P_2/P_1)$.  ]
% \color{black}
After that, it
continues to undergo  MMR-driven oscillations, but now around the other stable fixed point, with a roughly constant free eccentricity. %(eq. \ref{e:ec2}). 

%The jump magnitude is as described by eq. \refnew{eq:delta}, provided 
% $\dot{a}$ is sufficiently small.

The kick in $P_2/P_1$ is of order the width of the MMR, which, for an equal-mass pair, is approximated as
\citep{batyginadams},
% in particular, eq. 12 of batygin+adams, they refer back to batygin'15, but can't find it explicitly there...
\be
\Delta \left({{P_2}\over{P_1}}\right)_{\rm MMR} \approx 5 \left[ {{\sqrt{p+1} (M_{p1}+M_{p2})}\over{M_*}}\right]^{2/3}\, ,
\label{eq:delta}
\ee
%\x{\delta\approx 0.01 \left( {M_1\over 8M_\oplus} \right)^{2/3}\, ,
 and is $= 0.0095 (M_p/8M_\oplus)^{2/3}$ for the 3:2 MMR, 
 consistent with what is seen in the figure (1.505-1.4955=0.0095).
 For the 2:1 MMR, the kick is
  $0.0082 (M_p/8M_\oplus)^{2/3}$, as we have also verified with  REBOUND integrations.

The kick in $e_{21}$
may be approximated as follows.
From Fig. \ref{fig:simplepush}, we
see that immediately after the blue curve crosses the unstable fixed point, it hugs
a particular constant-$B$ curve. 
Given the kick in $\Delta\left({{P_2}\over{P_1}}\right)_{\rm MMR}$ along that constant-$B$ curve, we use the constancy of $B$ (Eq. \ref{eq:brou}) to find the corresponding kick in $e_{21}$, as
%follows from Eq. \ref{eq:brou}, which implies
%\be
%\Delta(e_{21}^2)={2\over 3p}\left({p\over p+1} \right)^{5/3}\Delta\left(  {P_2\over P_1}\right)_{\rm MMR}
%\approx 
%{10\over 3}{p^{5/3}\over (p+1)^{4/3}} 
%\left[ { M_{p1}+M_{p2}\over{M_*}}\right]^{2/3}
%\ee 
\be
\Delta (e_{21})_{\rm MMR}=\left({2\over 3p}\left({p\over p+1} \right)^{5/3}\Delta\left(  {P_2\over P_1}\right)_{\rm MMR}\right)^{1/2}
\approx 
1.8{p^{1/3}\over (p+1)^{2/3}} 
\left[ { M_{p1}+M_{p2}\over{M_*}}\right]^{1/3} \ .
\ee 
This corresponds to the net kick, starting from $e_{21}=0$ far from resonance. 
The above expression evaluates to
 $\Delta (e_{21})_{\rm MMR}=0.040(M_p/8M_\oplus)^{1/3}$ for the 3:2 MMR, consistent with the figure, and is 4\% larger than that for the 2:1 MMR.

\end{document}